\documentclass{aa}

\usepackage[varg]{txfonts}
\usepackage{amssymb}
\usepackage{epsfig}
\usepackage{graphics}
\usepackage{amsmath}
\usepackage{color}
\usepackage{natbib}

\bibpunct{(}{)}{;}{a}{}{,}

\newcommand{\be}{\begin{equation}}
\newcommand{\ee}{\end{equation}}
\newcommand{\bea}{\begin{align}\begin{split}}
\newcommand{\eea}{\end{split}\end{align}}

\newcommand{\Msun}{\ensuremath{\mathrm{M}_{\sun}}}

\begin{document}
\title{Equation of state constraints for the cold dense matter inside neutron stars using the cooling tail method}

\author{J.\,N\"attil\"a\inst{1}
  \and A.\,W.\,Steiner\inst{2}
  \and J.\,J.\,E.\,Kajava\inst{3}
  \and V.\,F.\,Suleimanov\inst{4,5}
  \and J.\,Poutanen\inst{1,6}
}

\institute{Tuorla Observatory, Department of Physics and Astronomy, University of Turku, V\"ais\"al\"antie 20, FI-21500 Piikki\"o, Finland \email{joonas.a.nattila@utu.fi}
  \and Department of Physics and Astronomy, University of Tennessee, Knoxville, Tennessee 37996, USA
  \and European Space Astronomy Centre (ESA/ESAC), Science Operations Department, 28691 Villanueva de la Ca\~nada, Madrid, Spain
  \and Institut f\"ur Astronomie und Astrophysik, Kepler Centre for Astro and Particle Physics, Universit\"at T\"ubingen, Sand 1, D-72076 T\"ubingen, Germany
  \and Astronomy Department, Kazan (Volga region) Federal University, Kremlyovskaya str. 18, 420008 Kazan, Russia
  \and Nordita, KTH Royal Institute of Technology and Stockholm University, Roslagstullsbacken 23, SE-10691 Stockholm, Sweden
}

\date{Received XXX / Accepted XXX}

\abstract{
The cooling phase of thermonuclear (type-I) X-ray bursts can be used to constrain the neutron star (NS) compactness by comparing the observed cooling tracks of bursts to accurate theoretical atmosphere model calculations.
By applying the so-called cooling tail method, where the information from the whole cooling track is used, we constrain the mass, radius, and distance for three different NSs in low-mass X-ray binaries 4U 1702$-$429, 4U 1724$-$307, and SAX J1810.8$-$260.
Care is taken to only use the hard state bursts where it is thought that only the NS surface alone is emitting.
We then utilize a Markov chain Monte Carlo algorithm within a Bayesian framework to obtain a parameterized equation of state (EoS) of cold dense matter from our initial mass and radius constraints.
This allows us to set limits on various nuclear parameters and to constrain an empirical pressure-density relation for the dense matter.
Our predicted EoS results in NS radius between $10.5-12.8~\mathrm{km}$ ($95\%$ confidence limits) for a mass of $1.4~\Msun$ depending slightly on the assumed composition.
Due to systematic errors and uncertainty in the composition these results should be interpreted as lower limits for the radius.
}


\titlerunning{Equation of state constraints for neutron stars using the cooling tail method}

\maketitle

\section{Introduction}\label{sec:intro}

The equation of state (EoS) of the cold dense matter inside neutron stars (NS) has remained a mystery for decades.
Experiments on Earth and theoretical many-body calculations have constrained the pressure-density relation of matter near the nuclear saturation densities.
Recently, progress has also been made in measuring the NS radii \citep[for a review, see][]{Miller13, O13, SPK15} that allow us to constrain the behavior of the EoS at higher densities by inverting the Tolman-Volkoff-Oppenheimer \citep[TOV;][]{Tolman39, OV39} structure equations.
Furthermore, these measurements probe the phase diagram of dense quantum chromodynamics at lower temperatures and higher baryon densities than the measurements of, for example, ultrarelativistic heavy-ion collisions inside earthly laboratories \citep[see, e.g.,][]{Lattimer12ARNPS}.

One of the most promising candidates for obtaining accurate astrophysical mass$-$radius ($M-R$) measurements has been the thermal emission originating from NS surface layers.
One possibility is to use the cooling of NS surface during type-I X-ray bursts from low-mass X-ray binary (LMXB) systems, where the cooling tail is shown to follow theoretical model predictions surprisingly well \citep{PNK14}.
In these systems, the NS is accompanied by a lighter, usually a main sequence or evolved late-type, star that fills its Roche lobe and transfers material through accretion disk onto the NS.
After accumulating enough material, the fuel is rapidly burned in a thermonuclear explosion occurring below the surface in the NS ocean.
Some of these bursts can be so energetic that the Eddington limit is reached, causing the entire NS photosphere to expand.
These photospheric radius expansion (PRE) bursts can then be used to obtain mass and radius measurements by comparing the cooling tail of the burst to accurate theoretical predictions \citep[for early work, see, e.g.,][]{Damen90,vP90,LvPT93}.

Recent studies \citep{SPRW11, PNK14, KNL14} have demonstrated that the X-ray burst cooling properties heavily depend on the accretion rate and spectral state of the source.
The key finding was that care must be taken to select only those bursts that show ``passive cooling", i.e. the ones that occur at the hard spectral state and at small accretion rate, where the extra heating from the in-falling material appears to be negligible.
\cite{KNL14} also showed that the evolution of the blackbody normalization can be used as a trace to pin down the passively cooling bursts used in the $M-R$ measurements.
The soft state bursts, on the other hand, show only weak or completely non-existent evolution of the normalization that is in contradiction with the theoretical atmosphere model predictions.

  In addition to only using the passively cooling bursts, it is possible to improve the analysis by using the information from the whole cooling tail by applying the so-called ``cooling tail method''.
In this recently developed method, the observed cooling track in the blackbody normalization $K$ vs. the flux $F$ (or rather $K^{-1/4}$ vs. $F$) plane is compared to the theoretical model evolution of the color-correction factor versus the luminosity (in units of the Eddington), $f_{\mathrm{c}} - L/L_{\mathrm{Edd}}$, that is the so-called color-correction curve.
By comparing the whole cooling track to the models (with variable $f_{\mathrm{c}}$) -- in contrast for example to the ``touchdown method'' where $f_{\mathrm{c}}$ is assumed to be constant -- we can infer more robust constraints from the data.
This also allows us to circumvent the problematic issue of deciding where the Eddington limit is reached and when the photospheric radius coincides with the NS radius \citep[see, e.g.,][]{SLB10}.
In fact, because it is the curvature of the evolving color-correction factor that is used to constrain the Eddington flux, the method is valid even for bursts that do not reach the Eddington limit \citep[see, e.g.,][]{ZCG12}.
So far, no work exists where the cooling tail method would have been applied with this kind of strict hard-state burst selection criteria for many sources simultaneously.
Therefore, we now pay special attention to the burst selection and choose only the most well-behaved hard-state PRE bursts for our analysis.
We then use these bursts to constrain the mass and radii of three different NSs by applying the cooling tail method to them.

Using these constraints we can then also go one step further and address the issue of unknown EoS of the cold dense matter inside neutron stars.
To do this, we use Bayesian inference to derive empirical pressure-density and mass-radius relations based on our burst results.
Using a Markov Chain Monte Carlo (MCMC) algorithm, we jointly fit the three $M-R$ constraints from each source allowing us to put astrophysical constraints on some of the nuclear physics parameters, such as the symmetry energy $\mathcal{S}$ and the pressure of neutron-rich matter at the saturation density $\mathcal{L}$ \citep{LS14b}.
In addition, the combination of the cooling tail observations and parameterized EoS allows us to make more accurate mass and radius measurements for each of the sources, indicating a new way of probing individual NS characteristics.

The paper is structured as follows.
In Section \ref{sect:data}, we present the methods used for the data reduction of the bursts.
Then, in Section \ref{sect:ct}, we use this data to obtain separate mass, radius, and distance constraints for the three sources in our sample.
In the second part of the paper, in Section \ref{sect:eos}, we use Bayesian analysis to obtain the parameterized EoS.
Finally, in the last Section \ref{sect:disc}, we discuss the constraints and compare our results to the previously made measurements.

\section{Data}\label{sect:data}

\begin{table*}
\caption{X-ray bursts used in the $M$-$R$ analysis.}
\centering
\begin{small}
\begin{tabular}[c]{l c c c c}
\hline
\hline  Source &   $N_{\mathrm{H}}$       & obsid & Date & $K_{\rm td2}/K_{\rm td}$$^a$\\
         &   ($10^{22}$ cm$^{-2}$) &       & (MJD) & \\
\hline
4U 1702$-$429        & 1.87$^b$ & 50025-01-01-00 & 51781.333039 & 2.2 \\
                   &          & 80033-01-01-08 & 52957.629763 & 2.0 \\
                   &          & 80033-01-19-04 & 53211.964665 & 2.2 \\
                   &          & 80033-01-20-02 & 53212.794286 & 2.1 \\
                   &          & 80033-01-21-00 & 53311.806086 & 2.1 \\
\hline
4U 1724$-$307        & 0.78$^c$ & 93044-06-04-00 & 54526.679905 & 2.3 \\
\hline
SAX J1810.8$-$2609   & 0.35$^d$ & 93044-02-04-00 & 54325.894492 & 3.1 \\
\hline
\end{tabular}
\end{small}
\label{tab:bursts}
\begin{center}
  {\small{
    $^a$ Ratio of the blackbody normalizations at half-touchdown flux and at the touchdown \citep[see][]{KNL14,PNK14}, \\
    $^b$ \cite{WGP13}, 
    $^c$ \cite{Kuulkers03},
    $^d$ \cite{Natalucci2000}
}}
   \end{center}
\end{table*}

\begin{figure}
\centering
\includegraphics[width=8.5cm]{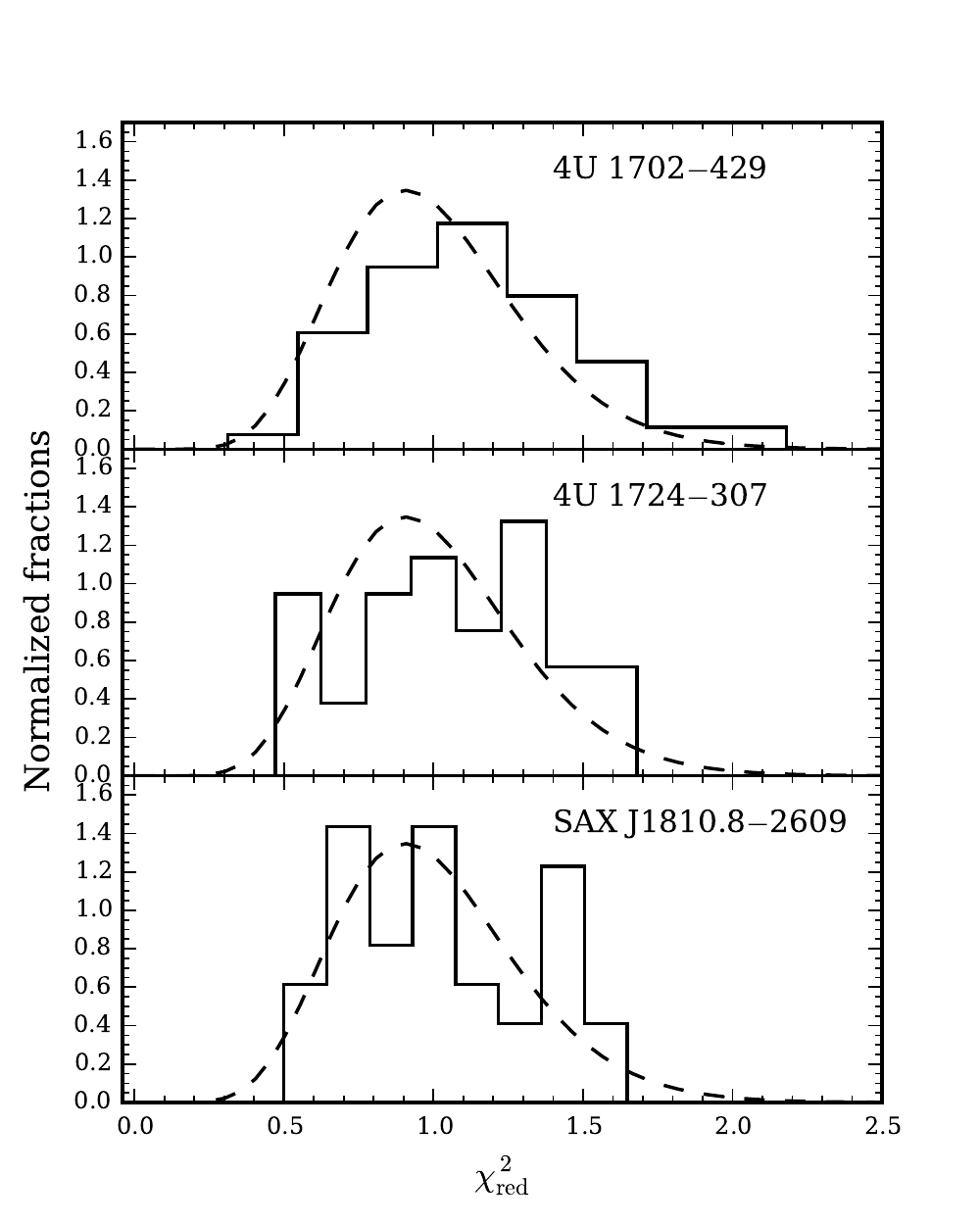}
\caption{\label{fig:chis}
  Reduced $\chi^2$ distributions for the black body spectral fits consisting of the points used in the cooling tail analysis.
  The dashed curve shows the theoretical expected $\chi^2$ distribution.
  Both obtained and theoretical distribution are normalized so that the encapsulated area of the curves is unity.
  }
\end{figure}

For our analysis we have used the data from the PCA \citep{JMR06} instrument on board of the {\it Rossi X-ray Timing Explorer (RXTE)} satellite.
Our sample consists of three neutron stars: 4U 1702$-$429, 4U 1724$-$307, and SAX J1810.8$-$2609.
These sources were selected because they have been known to exhibit PRE bursts in the hard state \citep[i.e. at low accretion rate where the NS is thought to cool passively without any external heating; ][]{KNL14}.
They also show the most robust evolution of the normalization down to very low luminosities enabling us to use them as clear examples of bursts for which the cooling tail method can be applied to.
  These bursts are visually selected, based on their evolution of normalization, to have a good agreement with the model used.
We omit one already analyzed long burst from 4U 1724$-$307 \citep{SPRW11, SPW11} as there is evidence that this particular burst might have a high metallicity content in the atmosphere (see Appendix \ref{sect:1724old}).

{\it RXTE}/PCA data were analyzed with the help of {\sc heasoft} package (version 6.16) and response matrices were generated using {\sc pcarsp} (11.7) task of this package.
All of the data were fitted using {\sc xspec 12.8.1} package \citep{Arn96} where the recommended systematic error of 0.5\% was added to the spectra \citep{JMR06}.
We identified the X-ray bursts using a similar method as in \citet{GMH08}.
The time-resolved spectra for the bursts were extracted using an initial integration time of 0.25~s.
In order to maintain approximately the same signal-to-noise ratio the integration time was doubled every time the count rate decreased by a factor of $\sqrt{2}$.
The exposures were dead-time corrected following the approach recommended by the instrument team.\footnote{\texttt{http://heasarc.gsfc.nasa.gov/docs/xte/recipes/ \\pca\_deadtime.html}}
The correction resulted in a roughly 10--15 per cent increase in the peak flux, with the difference decreasing quickly as the observed count rate declined.
A standard method of removing a 16~s spectrum taken from prior to the burst was used to account for the possible background emission \citep[][and references therein]{Kuulkers02}.
This standard method assumes the background (i.e. mainly the persistent emission) to be constant during the burst, even though this might not be the case.
The changes are, however, not significant in the cooling phase \citep[see Fig.~6 in ][]{WGP13}.
The differences in burst characteristics with and without this background subtraction was also checked and found to be negligible at least at high burst fluxes (i.e. at fluxes larger than $20$ per cent of the peak flux).
These deadtime-corrected spectra were then fitted with a blackbody model multiplied by an interstellar absorption model with constant hydrogen column density $N_{\mathrm{H}}$ (value obtained from the literature, see Table \ref{tab:bursts}).
The best-fit parameters are the color temperature $T_\mathrm{c}$ and the normalization constant $K\equiv(R_{\mathrm{bb}} [{\rm km}]/D_{10})^2$, where $D_{10} \!=\! D/10$ kpc.
From the corresponding $\chi^2$ distributions (see Fig.~\ref{fig:chis}) of each source we also conclude that the data is sufficiently well described by the blackbody model.
It should also be noted that the theoretical atmosphere model spectra cannot be perfectly fit by a (diluted) blackbody model either \citep{SPW11,SPW12}, so in reality we do not even expect the observed $\chi^2$ distribution to be close to ideal.
The bolometric flux was estimated using the \verb|cflux|-model in the range 0.01 -- 200 keV.
All error limits were obtained using \verb|error| -task in {\sc xspec}.

Some of these bursts show typical characteristics of a PRE:
A peak in the normalization after a few seconds of the ignition.
The evolution of the observed temperature should also show the characteristic double-peaked structure, arising from the cooling of the photosphere when it expands and the subsequent heating when it collapses back toward the surface due to the changing radiation pressure.
The aforementioned signs of the expansion also indicate that the flux has reached (or exceeded) the Eddington limit during the burst.
The moment when the atmosphere collapses back to the NS surface --- i.e. the normalization reaches its minimum value $K_{\mathrm{td}}$ and the temperature its second peak ---  is defined as the touchdown.
This also marks the beginning of the cooling phase where the subsequent evolution is dependent on the spectral state of the source, i.e. if the burst occurred in the hard state or in the soft state.
In the hard state the normalization rises to a nearly constant level while the flux and the temperature continue to decrease for the rest of the burst.
This increase of normalization is due to the changing color-correction factor $f_{\mathrm{c}}$ as we approximate the emerging spectrum with a diluted blackbody model as
\be
F_E \approx \frac{1}{f_{\mathrm{c}}^4} B_E (f_{\mathrm{c}} T_{\mathrm{eff}}),
\ee
where $B_E$ is the blackbody function and $T_{\mathrm{eff}}$ is the effective temperature that is connected to the observed blackbody color temperature as $T_{\mathrm{c}} = f_{\mathrm{c}} T_{\mathrm{eff}}(1+z)^{-1}$ , where $1+z = (1-2GM/Rc^2)^{-1/2}$ is the redshift.
We stress here that the decrease in the color-correction factor during the cooling is a feature predicted by numerous atmosphere model computations \citep{London86,LST86,SPW11, SPW12}.
Consequently, the decrease of the color correction then leads into an increase in the (observed) normalization because \citep{Penninx89, vP90}
\be\label{eq:acon}
K^{-1/4} = f_{\mathrm{c}} A,\phantom{AAAA} A = (R_{\infty} [\mathrm{km}]/ D_{10})^{-1/2},
\ee
where $R_{\infty} = R(1+z)$ is the apparent NS radius.
On the other hand, for the soft-state bursts the normalization is nearly constant, contrary to the theory.%
\footnote{One possible interpretation is that in the soft state the accretion disk continues all the way down to the NS surface forming a spreading/boundary layer.
  A combination of emission from a partly visible NS and from the spreading/boundary layer itself can then create time-evolving spectra that appear to have almost constant color-correction factor \citep{SulP06}.}
It is also crucial to notice that the value of the normalization in the tail of the soft-state bursts is different from what is observed for the hard-state bursts.
In addition, the touchdown flux can also vary%
\footnote{In the soft state the inner disk may act as a mirror reflecting part of the burst emission, therefore boosting the observed flux \citep{LS85}.
} \citep[see Fig.~1 in][]{KNL14}.
Because of these differences, the burst selection becomes extremely important as our model assumptions are only valid if the NS surface alone is emitting that seems to be valid only in the hard state \citep[see][ for more information about the soft vs. hard-state burst selection]{PNK14, KNL14}.
The increasing emission area during the PRE phase (i.e. increase in the blackbody normalization $K$) before the touchdown is mostly related to the increase of the photospheric radius. 
In the cooling tail, one believes that the evolution of $K$ happens just because of varying $f_{\rm c}$ with the constant actual photospheric radius equal to the NS radius. 
Therefore, the ratio of the normalizations at the expansion phase and the cooling tail is 
\be
\frac{K_{\mathrm{e}}} {K_{\mathrm{t}}} = 
\left( \frac{f_{\mathrm{c,t}}} {f_{\mathrm{c,e}}} \right)^4 
\left( \frac{(1+z_{\mathrm{e}})R_\mathrm{e}}  {(1+z_{\mathrm{t}})R_{\mathrm{t}}} \right)^2,
\ee
where  indices $\mathrm{e}$ and $\mathrm{t}$ refer to the expansion and tail, respectively.  
By taking $f_{\mathrm{c, t}} \approx 1.4$ in the tail and $f_{\mathrm{c, e}} \gtrsim 2$ during the expansion \citep{PSZ91, SPW12} and demanding that $R_\mathrm{e}>R_\mathrm{t}$ (which also means $(1+z_{\mathrm{e}})R_\mathrm{e} > (1+z_{\mathrm{t}})R_{\mathrm{t}}$ for $R > \frac{3}{2} \frac{2GM}{c^2}$), 
we end up with a simple PRE condition
\be\label{eq:PRE}
\frac{K_{\mathrm{e}}}{K_{\mathrm{t}}} \gtrsim \frac{1}{4}. 
\ee
When the normalizations are equal $K_{\mathrm{e}}=K_{\mathrm{t}}$, we get $R_\mathrm{e} \gtrsim 2\,R_\mathrm{t}$ (note also that 
$z_{\mathrm{t}} > z_{\mathrm{e}}$). 
What is remarkable with this condition is that the observed ``expansion'', can be less than unity (compared to the tail) in order for the burst to have a PRE episode.
The PRE condition can be transformed into a requirement that the observed peak normalization at the expansion phase $K_{\mathrm{e}}$ must be larger than the normalization at the touchdown $K_{\mathrm{td}}$ as both of them should have similar values of the color-correction factors.
But this is equivalent to the standard criterion that $K$ should have a local minimum (at the touchdown).

\begin{figure*}
\centering
\includegraphics[width=18.5cm]{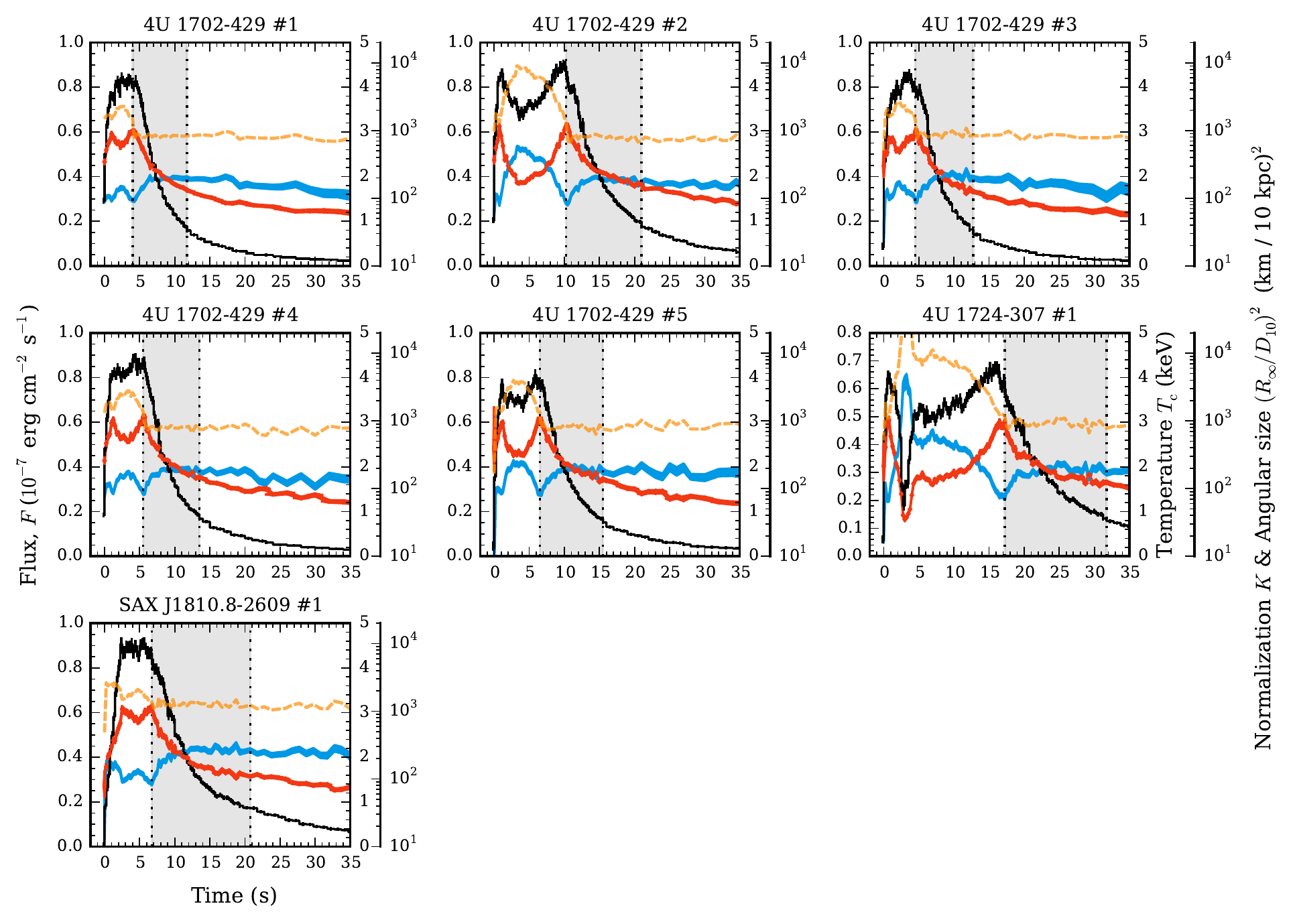}
\caption{\label{fig:bursts}
  Bolometric flux, temperature and blackbody normalization evolution during the hard-state PRE bursts.
  The black line shows the estimated bolometric flux (left-hand y-axis) in units of $10^{-7}$erg cm$^{-2}$ s$^{-1}$.
  The blue ribbon shows the 1$\sigma$ limits of the blackbody normalization (outer right-hand y-axis) in $(\mathrm{km}/{10\mathrm{kpc}})^2$.
  Similarly, the dashed orange line shows the color-corrected angular size $(R_{\infty}/D_{10})^2$ using the same axis.
  The red ribbon show the 1$\sigma$ errors for blackbody color temperature (inner right-hand y-axis) in keV.
  Highlighted gray area marks the region of the cooling tail used in the fitting procedures.
  }
\end{figure*}

Time-resolved spectral parameters for the bursts in our sample are presented in Fig.~\ref{fig:bursts}.
Additionally, we show the color-corrected angular size with the assumption $f_{\mathrm{c}} = 2$ for the evolution before the touchdown.
For the $f_{\mathrm{c}}$ values after the touchdown, we use the cooling tail model fits from Sect.~\ref{sect:ct} to correct for the varying color-correction factor.
Because of the new PRE criteria we choose to keep also the bursts that show only modest photospheric expansion in our sample (bursts $\#1$ and $\#3$ from 4U 1702$-$429 and burst $\#1$ from SAX J1810.8$-$2609).
Even though the expansion phase in these bursts is not very long, it is obvious from the Fig.~\ref{fig:bursts} that the subsequent cooling phase is still similar (compare, for example, the bursts $\#1$ and $\#2$ from 4U 1702$-$429).

\section{The Bayesian cooling tail method}\label{sect:ct}

For our mass and radius analysis we use the cooling tail method (see \citealt{SPRW11,SPW11} and Appendix A of \citealt{PNK14}).
With this method the information from the whole cooling track after the peak of the burst is used and the observed cooling is compared to the theoretical evolution predicted by passively cooling NS atmosphere models.
To relate the observed data and the individual masses and radii of the NSs, we use Bayesian analysis \citep[see also][]{OP15}.
Bayes' theorem can be presented simply as \citep[see, e.g.,][]{GS97}
\be
\Pr(\mathcal{M}|\mathcal{D}) = \frac{\Pr(\mathcal{D}|\mathcal{M})\Pr(\mathcal{M})}{\Pr(\mathcal{D})},
\ee
where $\Pr(\mathcal{M})$ is the prior probability of the model $\mathcal{M}$ without any additional information from the data $\mathcal{D}$, $\Pr(\mathcal{D})$ is the prior probability of the data $\mathcal{D}$, $\Pr(\mathcal{D}|\mathcal{M})$ is the conditional probability of the data $\mathcal{D}$ given the model $\mathcal{M}$ and $\Pr(\mathcal{M}|\mathcal{D})$ is the conditional probability of the model $\mathcal{M}$ given the data $\mathcal{D}$.
Here the last quantity $\Pr(\mathcal{M}|\mathcal{D})$ is what we want to obtain and it gives us the probability that a given model is correct given the data.
We can extend the Bayes theorem further by having many non-overlapping models $\mathcal{M}_i$, which exhaust the total model space $\mathcal{M}$.
Then the relation can be written as
\be
\Pr(\mathcal{M}_i|\mathcal{D}) = \frac{\Pr(\mathcal{D}|\mathcal{M}_i)\Pr(\mathcal{M}_i)}{\sum_j \Pr(\mathcal{D}|\mathcal{M}_j) \Pr(\mathcal{M}_j)}.
\ee
In the cooling tail method the model space consists of four parameters:
mass $M$ and radius $R$ of the NS, hydrogen mass fraction $X$ in the atmosphere, and the distance $D$ to the source.
For the distance $D$ a uniform flat prior distribution is assumed without any restrictions.
Similarly, a uniform 2-dimensional prior distribution is assumed for $(M,R)$ space.
We also take into account the causality-requirement, $R > 2.824 GM/c^2$ \citep{Lattimer12ARNPS} and limit the mass to lie between $0.8\Msun < M < 2.5\Msun$.
For the hydrogen fraction $X$, a Gaussian prior distribution is used and is discussed in more detail later on.

The model parameters can be combined into two new parameters, related more closely to the color-correction curve fitting.
The first one is the Eddington flux
\be\label{eq:fedd}
F_{\mathrm{Edd}} = \frac{G M c}{D^2 \kappa_{\mathrm{e}} (1+z)},
\ee
where $\kappa_{\mathrm{e}} = 0.2(1+X)~\mathrm{cm}^2\,\mathrm{g}^{-1}$. 
The second parameter is related to the apparent (non-color corrected) angular size \eqref{eq:acon}.
These parameters then relate our observed flux to the (relative) luminosity as $F/F_{\mathrm{Edd}} \propto L/L_{\mathrm{Edd}}$ (where $L_{\mathrm{Edd}}$ is the Eddington luminosity) and observed blackbody normalization to the color-correction as $K^{-1/4}=f_{\mathrm{c}} A$.
We also note here that it is possible to assume uniform priors for $F_{\mathrm{Edd}}$ and $A$ (in contrast to assuming uniform flat distribution for $(M,R)$ space; see Appendix \ref{sect:AppendixB}) as was done previously by \citet{SPW11, SPW12, PNK14}.

As our actual model, we use the recently computed hot neutron star atmosphere models \citep{SPW12} that account for the Klein-Nishina reduction of the electron scattering opacity using an exact relativistic Compton-scattering kernel \citep{PS96}.
These models give us the color-correction as a function of relative luminosity, $f_{\mathrm{c}}(\ell \equiv L/L_{\mathrm{Edd}})$.
The model uncertainty is taken into account by considering a boxcar distribution of a width of $(1 \pm \epsilon) \times f_{\mathrm{c}}$ (where $\epsilon=0.03$) centered around the ``real'' value (see \citet{SPW11, SPW12} for a discussion of model uncertainties).
Compositions considered are a pure helium (He) and a solar composition of H and He with sub-solar metal abundance of $Z = 0.01 Z_{\sun}$ (SolA001).
It seems that $Z \!<\! 0.1 Z_{\sun}$ in the surface layers of the NS, because in the opposite case the atmosphere model predicts a drop of around $~20\%$ in the $f_{\rm c}$ (and correspondingly in  $K^{-1/4}$) at $F\sim 0.3 F_{\mathrm{Edd}}$ \citep{SPW11,SPW12}, which is not observed. 
Alternatively, the absence of the drop in the low luminosities might be due to an extra heating because of the accretion that starts again after the burst.
In any case, owing to these uncertainties in the low luminosity regime, we neglect this area from the fit and consider only the regime where $F > 0.2 F_{\mathrm{td}}$, where $F_{\mathrm{td}}$ is the flux at the touchdown point.

We also relax the assumption of having a fixed hydrogen mass fraction $X$ ($X=0.738$ for SolA001 and $X=0$ for He models) and use Gaussian priors around the exact values with one sigma tails of $0.05$.
Both compositions are tested for each source and the physically more justified value is selected.
Note also that this selection is easy as wrong composition gives $R \lesssim 6~\mathrm{km}$ or $R \gtrsim 18~\mathrm{km}$.
Strictly speaking this should be taken into account by using atmosphere models that are computed with the corresponding hydrogen fractions but the models (i.e. color-correction factors $f_{\mathrm{c}}$) depend so weakly on this value (as our one sigma limits were $X = 0^{+0.05}_{-0.0}$ or $X=0.738 \pm 0.05$) that it is possible to neglect the effect that this has on the $F_{\mathrm{Edd}}$ and $A$ \citep[see, e.g.,][ where the difference is relatively small even for $X=0$ compared to $X=1$]{SPW12}.
For the $M$, $R$, and $D$ this, however, has some non-negligible effect that introduces a small scatter of about $5$ per cent to the posterior distributions around the ``exact'' value.
The uncertainty in the hydrogen fraction is also backed up by physical arguments because for hydrogen-poor companions (in the case of $X=0$, i.e. He models) the evolutionary models do not exclude the possibility of the envelope containing some fraction of H \citep[$X \lesssim 0.1$;][]{PRP02}.
The value of solar ratio of H/He is relatively accurately measured but here the uncertainties are possible and due to the possible stratification on top of the NS and/or because of the light ashes from the previous bursts that may stratify and accumulate to the surface.
In the end, however, one should remember that the value of the hydrogen mass fraction for each model is still just a model assumption, made on some basis.
By selecting a Gaussian prior around the presumed value we do weaken the effect that this selection has but we are unable to remove it completely.
If no assumption for the hydrogen mass fraction would be made, we could not infer the radius at all.
Reassuringly, however, the end results do seem to gather around similar radii which means that our assumed $X$ values were close to the real values.

In our method the data $\mathcal{D}$ is constructed as a set of $N$ points $(F_i, K_i^{-1/4})$ obtained from the blackbody fits, starting from the touchdown ($i=1$) and continuing down to $0.2 F_{\mathrm{td}}$ ($i=N$).
  The lower-limit here is selected so that we can maximize the available data (in contrast to $1/\mathrm{e}~F_{\mathrm{td}}$ used in previous work) as the theoretical models nicely follows the data. 
Below the $0.2 F_{\mathrm{td}}$ limit the background emission can start to play too important role so we choose to leave it out even though some of the bursts might follow the model even beyond this.
These data points are then transformed into 2-dimensional probability density distributions $\mathcal{D}_i(F, K^{-1/4})$ by assuming a Gaussian measurement error model.
Next we implicitly assume that all of the data distributions $\mathcal{D}_i$ are independent of each other and also independent of the model assumptions and prior distributions.
We then assume that the conditional probability of the data given the model, $\Pr(\mathcal{D}|\mathcal{M})$, is proportional to the product over every individual probability $\Pr(\mathcal{D}_i|\mathcal{M})$ 
\be
\Pr(\mathcal{D}|\mathcal{M}) \propto \prod_{i=1,\ldots,N_S} \Pr(\mathcal{D}_i|\mathcal{M}).
\ee
Each separate probability $\Pr(\mathcal{D}_i|\mathcal{M})$ is assumed to be proportional to the path-integral evaluated through the 2-dimensional ``data space'', $(F, K^{-1/4})$, along the color-correction curve as
\begin{align}\begin{split}\label{eq:path-int}
    &\Pr[\mathcal{D}_i|\mathcal{M}(M, R, D, X)] \\
    &\propto \frac{1}{\mathcal{N}}\int_{f_{\mathrm{c,lo}}}^{f_{\mathrm{c,hi}}} \mathrm{d}f_{\mathrm{c}}(\ell)
    \int_{\mathcal{F}_{\mathrm{c}}} \mathcal{D}_i(F, K^{-1/4}) \, J\left(\frac{\ell, f_{\mathrm{c}}}{F, K^{-1/4}} \right) \mathrm{d}s
\end{split}\end{align}
where $f_{\mathrm{c, lo, hi}} = (1 \pm \epsilon) \times f_{\mathrm{c}}(\ell)$ are the lower and upper limits of the prior boxcar distribution of the color-correction $f_{\mathrm{c}}$ evaluated at relative luminosity $\ell$ and where $\epsilon = 0.03$, $\mathcal{F}_{\mathrm{c}} = \mathcal{F}_{\mathrm{c}}(F_{\mathrm{Edd}}, A)$ is the color-correction curve in $(\ell, f_{\mathrm{c}})$ space, $J\left(\frac{\ell, f_{\mathrm{c}}}{F, K^{-1/4}} \right)$ is the Jacobian transforming the path from the ``model space'' to the observed ``data space'' (using equations \eqref{eq:acon} and \eqref{eq:fedd}), and $\mathrm{d}s$ is the line element of $\mathcal{F}_{\mathrm{c}}$.
The path-integral is also area-normalized (or length-normalized if $\epsilon=0$) with the factor $\mathcal{N}$ that is defined as the aforementioned integral \eqref{eq:path-int} without the data $\mathcal{D}$.
This normalization takes into account the variable maximum $\ell$ that evolves as a function of $\log g$.
We also note that the presented Bayesian path-integral formalism is related to the 2-dimensional frequentist formulation of the normalized minimum distance \citep[see eq. (2) in ][]{PNK14}.

In Bayesian interference, model parameters are determined using a marginal estimation where the posterior probability for a model parameter $p_j$ is given by
\begin{align}\begin{split}\label{eq:Bayesf}
& \Pr[p_j|\mathcal{D}](p_j) = \frac{1}{V} \int \Pr[\mathcal{D}|\mathcal{M}]\\
& \times \mathrm{d}p_1 \mathrm{d}p_2\ldots \mathrm{d}p_{j-1} \mathrm{d}p_{j+1}\ldots \mathrm{d}p_{N_P},
\end{split}\end{align}
where $N_P = 4$ (corresponding to $M$, $R$, $D$, and $X$) and
\be
V = \int \Pr[\mathcal{D}|\mathcal{M}] \Pr[\mathcal{M}]\mathrm{d}^N \mathcal{M},
\ee
without the model priors that determine the integration limits.
Then the one-dimensional function $\Pr[p_j|\mathcal{D}](p_j)$ represents the probability that the $j$th parameter will take a particular value given the observational data $\mathcal{D}$.

\begin{table*}
\caption{Results of the Bayesian cooling tail analysis}
\centering
\begin{small}
\begin{tabular}[c]{l c c c c}
\hline
\hline
  Source &  Model &  $F_{\mathrm{Edd}}$  & $A$ & $D$ \\
  &                   & ($10^{-7}$ erg cm$^{-2}$ s$^{-1}$) & ([km/10~kpc]$^{-1/2}$)  &  (kpc) \\[3pt]
\hline
4U 1702-429        & He & $0.80^{+0.01 ~(+0.02)}_{-0.02 ~(-0.02)}$ & $0.192^{+0.001 ~(+0.002)}_{-0.002 ~(-0.004)}$ &  $5.6^{+0.3 ~(+0.6)}_{-0.4 ~(-0.9)}$ \\[5pt]
4U 1724-307        & SolA001 & $0.58^{+0.02 ~(+0.04)}_{-0.02 ~(-0.03)}$ & $0.184^{+0.002 ~(+0.003)}_{-0.003 ~(-0.004)}$ &  $4.9^{+0.5 ~(+0.7)}_{-0.6 ~(-1.1)}$ \\[5pt]
SAX J1810.8-2609   & SolA001 & $0.79^{+0.02 ~(+0.03)}_{-0.02 ~(-0.03)}$ & $0.169^{+0.002 ~(+0.004)}_{-0.002 ~(-0.003)}$ &  $4.3^{+0.4 ~(+0.6)}_{-0.5 ~(-1.0)}$ \\[5pt]
\hline
\end{tabular}
\end{small}
\label{tab:MCMC}
\begin{center}
  {\small{
      Notes: Errors correspond to the $68\%$ and $95\%$ (in parentheses) confidence levels.\\
}}
   \end{center}
\end{table*}

\begin{figure}
\centering
\includegraphics[width=9.0cm]{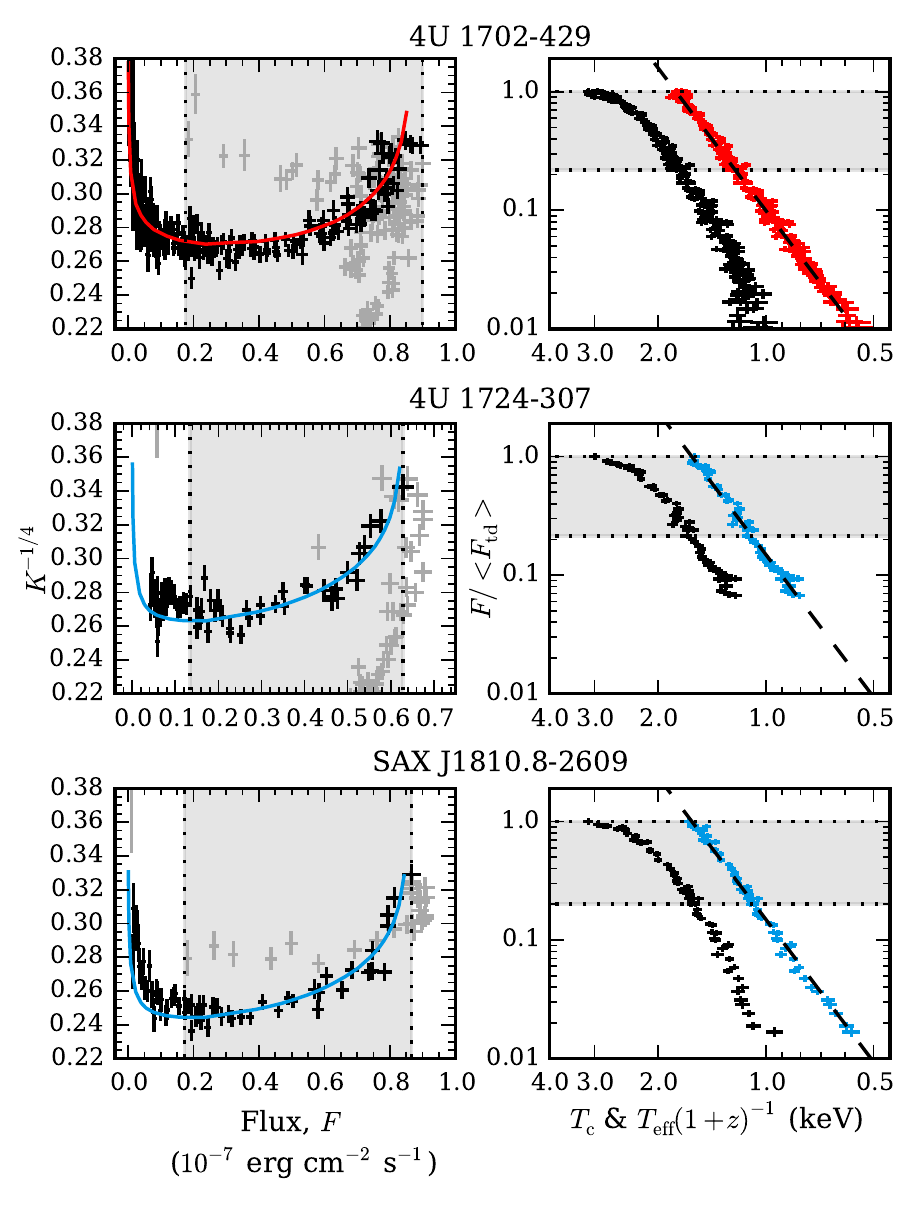}
\caption{\label{fig:cooling_tails}
  \textit{Left panel:} Combined cooling tail in the $F \propto L/L_{\mathrm{Edd}}$ vs $K^{-1/4} \propto f_{\mathrm{c}}$ plane with 1$\sigma$ error limits presented by crosses.
  Similarly, the gray crosses show the burst evolution before the touchdown.
  Best-fit theoretical atmosphere models are shown by the blue (SolA001) or red (He) curves.
  \textit{Right panel:} Temperature evolution of the bursts.
  Blackbody color temperature $T_{\mathrm{c}}$ is shown for each cooling tail with black crosses.
  Red (He) or blue (SolA001) crosses show the color-corrected temperatures $T_{\mathrm{eff}}(1+z)^{-1}$.
  The dashed lines show a powerlaw with an index of 4 corresponding to the $F \propto T^4$ relation.
  Highlighted gray area marks the region of the cooling tail used in the fitting procedures.
}
\end{figure}

\begin{figure}
\centering
\includegraphics[width=9cm]{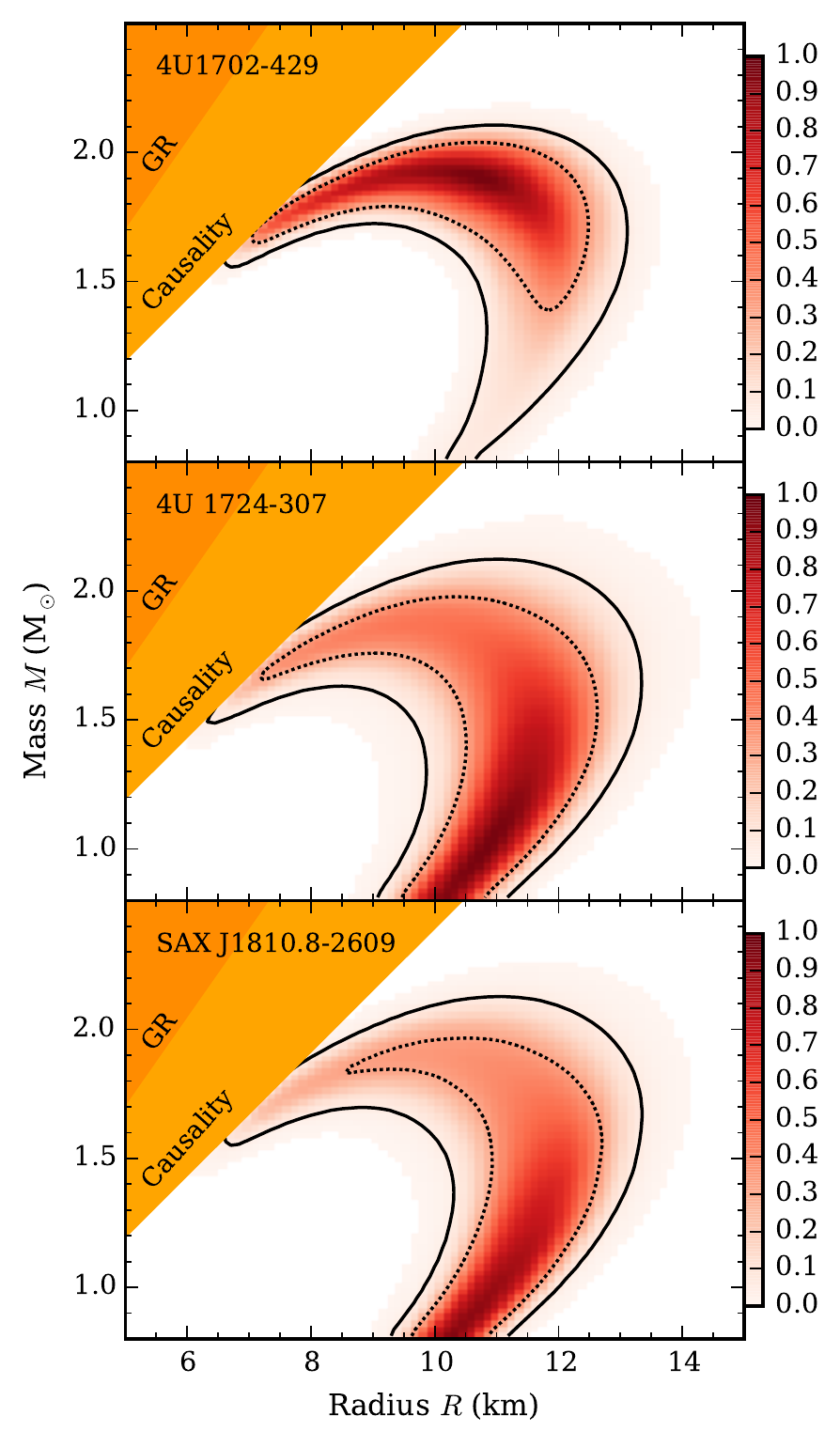}
\caption{\label{fig:MR}
  Mass-radius constraints for the sources from the hard state PRE bursts.
Constraints are shown by $68\%$ (dotted line) and $95\%$ (solid line) confidence level contours.
The upper-left region is excluded by constraints from the causality and general relativistic requirements \citep{HPY07,LP07}. 
}
\end{figure}

The best-fit atmosphere models are presented in Fig. \ref{fig:cooling_tails} (left panel).
In addition, the right panel of the figure depicts the observed color temperatures $T_{\mathrm{c}}$ and the corrected effective temperatures $T_{\mathrm{eff}}(1+z)^{-1}$ for a distant observer.
Here the temperature is seen to follow the $L \propto T_{\mathrm{eff}}^4$ law, i.e. showing passive cooling.

Corresponding best-fit values of the model parameters $F_{\mathrm{Edd}}$ and $A$ are listed in Table \ref{tab:MCMC} along with the $1\sigma$ and $2\sigma$ confidence limits of the posterior distributions.
After marginalizing over the radius $R$, mass $M$, and hydrogen mass fraction $X$ we get the posterior distribution for the distance $D$ too, that is also listed in Table \ref{tab:MCMC}. 
Fig. \ref{fig:MR} shows the 2-dimensional mass and radius probability posterior distributions of our analysis.
The obtained contours are arched and elongated along the curves of constant Eddington temperature%
\footnote{Eddington temperature formulated using the $F_{\mathrm{Edd}}$ and $A$ parameters is not strictly constant in the $M-R$ plane because $F_{\mathrm{Edd}}$ has a $\log g$ dependency (because of the dependence of $f_{\mathrm{c}}$ on $\log g$).
  This complication is introduced via the new models \citep{SPW12} that formally have super-Eddington luminosities due to the Klein-Nishina reduction of the effective cross-section.
We note that our new cooling tail formalism allows to take this into account as we use $M$ and $R$ as our parameters (instead of $F_{\mathrm{Edd}}$ and $A$).}
\be\label{eq:tedd}
T_{\rm Edd,\infty}  = \left(\frac{gc}{\sigma_{\rm SB} \kappa_{\rm e} }\right)^{1/4} \!\! \frac{1}{1+z} = 1.14\times 10^8\, A\,F_{-7}^{1/4}\ \mbox{K} , 
\ee 
where $g$ is the surface gravity, $\sigma_{\mathrm{SB}}$ is the Stefan-Boltzmann constant and $F_{-7} = F_{\mathrm{Edd}} / 10^{-7}~\mathrm{erg}\,\mathrm{cm}^{-2}\,\mathrm{s}^{-1}$.
The width of the contours are defined by the errors in $T_{\rm Edd,\infty}$ that originate from the uncertainty in $F_{\mathrm{Edd}}$, $A$, and $X$.
Our location on this curve is defined by the distance $D$.
Because the distance is free to vary in our analysis we end up with the aforementioned arched posteriors.

Contours from 4U 1724$-$307 and SAX J1810.8$-$2609 are seen to be almost identical with the radius constrained between about $11 - 13$ km (for a more strict lower mass limit of $M > 1.1 \Msun$) where the largest scatter is being produced by the unknown distance.
Both of these sources are also best explained by a solar-like composition with an almost zero metallicity (SolA001 model).
For 4U 1702$-$429 the radius is seen to be on the same range if mass $\lesssim 1.8~\Msun$ is assumed.
The model in this case consists of a pure helium composition.

The choice of pure helium composition causes the Bayesian model to favor larger mass.
This happens because $F_{\mathrm{Edd}} \propto M/(1+X)$, so when the hydrogen fraction increases/decreases (because of the possible uncertainty in the $X$ parameter) the change can be balanced by increasing/decreasing $M$ as well.
With the solar composition of H and He our $X$ priors are symmetrical and this effect is canceled out.
In the case of He models the lower-limit of $X=0$ makes the hydrogen fraction prior distribution asymmetrical and hence the bias for larger mass is present.
We note, however, that in addition to this bias there is a slight preference for the 4U 1702$-$429 to favor larger $\log g$ values and hence larger masses.
Similar effect is also present in the $M$ vs. $R$ posteriors because of our choice of flat distance prior.
As $F_{\mathrm{Edd}} \propto M/D^2$ we end up oversampling the small distance values of our flat prior that can then create a bias that favors smaller mass.
This effect is visible as an increased probability density around $M \lesssim 1.5$ for 4U 1724$-$307 and SAX J1810.8$-$2609.
Because of these model biases one should be careful not to give too much emphasis on the specific values of the $M$ vs. $R$ distributions presented in Fig \ref{fig:MR} but to focus more on the overall structure of the posteriors given by the confidence contours.

It is also possible to set some constraints to the unknown distance by using the cooling tail method.
From the values of $F_{\mathrm{Edd}}$ and $A$, inferred from the data, we can derive the maximum distance where we still have $M$ and $R$ solutions \citep[see Appendix A of][for the full set of equations]{PNK14} as
\begin{equation}\label{eq:dmax}
 D_{10} \le D_{10,\mathrm{max}} = \frac{1.77 \times 10^{-2}}{(1+X) A^2 F_{-7}}.
\end{equation}
On the other hand, our lower-limit of the mass prior distribution ($M_{\mathrm{min}} = 0.8~\Msun$) also sets a lower-limit to the distance when combined with $F_{\mathrm{Edd}}$ and $A$.
From these two constraints we are then able to put some limits on the distance to the NS too.

\section{EoS constraints}\label{sect:eos}
The final goal of the mass and radius measurements is to constrain the pressure-density relation of the cold dense matter.
Here we use these new mass and radius constraints from the three NSs to probe the EoS by applying a Bayesian analysis to the data.

Here the model space consists of EoS parameters $p_{i=1,\ldots,N_p}$ in addition to the values of neutron star masses $M_{i=1,\ldots,N_M}$ with a total dimensionality of our model space as $N=N_p + N_M$.
The total number of neutron stars in our sample is $N_M=3$ and the number of EoS parameters $N_p$ depends on our initial choice of the model (see Sect. \ref{sect:eos_param}).

The data $\mathcal{D}$ is now constructed as a set of $N_M$ probability distributions, $\mathcal{D}_i(M,R)$ in the $(M,R)$ plane obtained from the cooling tail posteriors presented in Section \ref{sect:ct}.
All of these distributions are normalized to unity by computing the integral
\be
\int_{M_{\mathrm{low}}}^{M_{\mathrm{high}}} \mathrm{d}M \int_{R_{\mathrm{low}}}^{R_{\mathrm{high}}} \mathrm{d}R \, \mathcal{\mathcal{D}}_i(M,R) = 1~\forall~i.
\ee
This normalization ensures that each source is treated equally in the analysis.
As integration limits we use the same constraints as in the cooling
tail method analysis where $M_{\mathrm{low}}=0.8~\Msun$, $M_{\mathrm{high}}=2.5~\Msun$, $R_{\mathrm{low}}=5~\mathrm{km}$ and $R_{\mathrm{high}}=18~\mathrm{km}$.

Next we again implicitly assume that all of the new data distributions $\mathcal{D}_i$ are independent of each other and also independent of the model assumptions and prior distributions.
We then assume that the conditional probability of the data given the model, $\Pr(\mathcal{D}|\mathcal{M})$, is proportional to the product over the probability distributions $\mathcal{\mathcal{D}}_i$ evaluated at some mass $M_i$ and radius (determined from the model) $R(M_i)$ so that
\begin{align}\begin{split}
    \Pr[\mathcal{D}|\mathcal{M}&(p_{1,\ldots,N_P}, M_{1,\ldots,N_M})] \\
    &\propto \prod_{i=1,\ldots,N_M} \mathcal{\mathcal{D}}_i(M,R) |_{M=M_i, R=R(M_i)}.
\end{split}\end{align}
For the model parameters $n_p + n_M$ we assume a uniform distribution (i.e. weakly informative physical priors) except for a few physical constraints described below.

In order to obtain all of the posterior probability distributions for the parameters we use the publicly available \texttt{bamr} code \citep{bamr}.
The TOV solver and data analysis routines were obtained from the O$_2$scl library \citep{o2scl}.
To solve the integral \eqref{eq:Bayesf} the code uses the Metropolis-Hastings algorithm to construct a Markov chain to simulate the distribution $\Pr[\mathcal{D}|\mathcal{M}(p_{1,\ldots,N_P}, M_{1,\ldots,N_M})]$.
For each point, an EoS parameter $p_i$ and the neutron star mass $M_i$ is generated.
Corresponding radius curve $R(M)$ (and radius $R_i$) is then obtained by solving the TOV equations.
From these three masses and radii, the weight function $\Pr[\mathcal{D}|\mathcal{M}]$ is computed and the point is either accepted or rejected according to the Metropolis algorithm.

\subsection{EoS parameterization}\label{sect:eos_param}

When building our EoS we follow the work by \cite{SGF15} and separate our density into three effective regimes:
the crust and regions below and above the nuclear saturation density $n_0$.
We assume nuclear binding energy of $E_{\mathrm{nuc}}(n_0) = -16$ MeV and saturation density of $0.16$ fm$^{-3}$, typical values obtained from \cite{KLM10}.
In mass densities these values correspond to about $2.7\times10^{14}~\mathrm{g}\,\mathrm{cm}^{-3}$.
The nuclear symmetry energy (the difference between energy per baryon of neutron matter and that of the nuclear matter)\footnote{Quartic terms are ignored, see \cite{Steiner06}.} is denoted as $\mathcal{S}(n_{\mathrm{B}})$, where $n_{\mathrm{B}}$ is the baryon number density and $\mathcal{S} \equiv \mathcal{S}(n_0)$.
The pressure of neutron-rich matter at the saturation density $n_0$ is denoted by $\mathcal{L} \equiv 3n_0\mathcal{S}'(n_0)$.
In addition, the nuclear masses and theoretical models imply a correlation between $\mathcal{L}$ and $\mathcal{S}$ \citep{LS14b} and thus we additionally restrict these parameters as $(9.17\mathcal{S} - 266~\mathrm{MeV}) < \mathcal{L} < (14.3\mathcal{S}-379~\mathrm{MeV})$ enclosing the aforementioned constraints.\footnote{More accurately speaking, the constraints originate from nuclear masses \citep{KLM10}, quantum Monte Carlo model \citep{GCR12}, chiral interactions \citep{TKH13}, and from isobaric analog states \citep{DL14}.}
The transition density from the crust to core is fixed to be at nuclear baryon density of $0.08~\mathrm{fm}^{-3}$.
We note that the crust model (and neither the fixed transition density) has almost no effect to the resulting radii as the results are much more dependent on the high density behavior of our EoS.

\begin{table}
\caption{Prior limits for the EoS parameters}
\centering
\begin{small}
\begin{tabular}[c]{l c c}
 \hline
 \hline
  Quantity &  Lower limit & Upper limit \\
  \hline
  \multicolumn{3}{c}{QMC parameters}\\
  $a$ (MeV) & 12.5 & 13.5 \\
  $\alpha$ & 0.47 & 0.53 \\
  $\mathcal{S}$ (MeV) & 29.5 & 36.1 \\
  $\mathcal{L}$ (MeV) & 30.0 & 70.0 \\
  \hline
  \multicolumn{3}{c}{Model A parameters}\\
  $n_1$ & 0.2 & 8.0 \\
  $\epsilon_1$ (MeV fm$^{-3}$) & 150 & 1600 \\
  $n_2$ & 0.2 & 8.0 \\  
  $\epsilon_2$ (MeV fm$^{-3}$) & 150 & 1600 \\
  $n_3$ & 0.2 & 8.0 \\
  \hline
  \multicolumn{3}{c}{Model C parameters}\\
  $\Delta P_1$ (MeV/fm$^{3}$) & 0 & 60 \\
  $\Delta P_2$ (MeV/fm$^{3}$) & 0 & 300 \\
  $\Delta P_3$ (MeV/fm$^{3}$) & 0 & 500 \\
  $\Delta P_4$ (MeV/fm$^{3}$) & 0 & 500 \\
\hline
\end{tabular}
\end{small}
\label{tab:param_priors}
\end{table}

\begin{table*}[!ht]
\caption{Most probable values and $68\%$ and $95\%$ confidence limits
  for the EoS parameters.}
\centering
\begin{footnotesize}
  \begin{tabular}[c]{l c c c c c}
    \hline
    \hline
  Quantity & $95\%$ lower limit & $68\%$ lower limit & Most probable value / median & $68\%$ upper limit & $95\%$ upper limit \\
  \hline
  \multicolumn{6}{c}{QMC parameters (with Model A)}\\
  $\mathcal{S}$ (MeV) & 29.6 & 30.4 & 32.2 & 33.3 & 35.0 \\
  $\mathcal{L}$ (MeV) & 32.1 & 42.1 & 54.9 & 67.7 & 69.4 \\
  \hline
  \multicolumn{6}{c}{Model A parameters}\\
  $n_1$ & 0.36 & 0.45 & 0.55 & 0.66 & 0.68 \\
  $\epsilon_1$ (MeV fm$^{-3}$) & 156 & 164 & 712 & 865 & 1020 \\
  $n_2$ & 0.25 & 0.25 & 0.47 & 4.80 & 7.55 \\
  $\epsilon_2$ (MeV fm$^{-3}$) & 531 & 794 & 1190 & 1510 & 1560 \\
  $n_3$ & 0.95 & 0.99 & 1.41 & 6.80 & 7.76 \\
  \hline
  \hline
  \multicolumn{6}{c}{QMC parameters (with Model C)}\\
  $\mathcal{S}$ (MeV) & 29.7 & 30.4 & 31.8 & 33.6 & 35.2 \\
  $\mathcal{L}$ (MeV) & 32.0 & 41.4 & 54.9 & 68.4 & 69.4 \\
  \hline
  \multicolumn{6}{c}{Model C parameters}\\
  $\Delta P_1$ (MeV/fm$^{3}$) & 5.0 & 9.9 & 15 & 23 & 31 \\
  $\Delta P_2$ (MeV/fm$^{3}$) & 59 & 122 & 176 & 194 & 195 \\
  $\Delta P_3$ (MeV/fm$^{3}$) & 44 & 186 & 345 & 386 & 390 \\
  $\Delta P_4$ (MeV/fm$^{3}$) & 12 & 26 & 199 & 372 & 385 \\
  \hline
 \end{tabular}
\end{footnotesize}
\label{tab:param_posteriors}
\begin{center}
  {\small{
      Notes: For the $\mathcal{L}$ and $\Delta P_4$ parameters we give the median value of the flat distribution between the $1\sigma$ limits.
}}
   \end{center}
\end{table*}

Near the nuclear saturation density, below the core we employ the ``Gandolfi-Carlson-Reddy'' (GCR) quantum Monte Carlo model \citep{GCR12} that takes into account the three-body forces between the particles in the high density matter.
The GCR results are accurately approximated by a two polytrope model given in terms of the energy $E$ for the neutron matter at some nucleon number density $n$ as
\be\label{eq:QMC_E}
E(n) = a \left(\frac{n}{n_0}\right)^{\alpha} + b \left(\frac{n}{n_0}\right)^{\beta} + m_n
\ee
with coefficients ($a$ and $b$) and exponents ($\alpha$ and $\beta$) constrained by QMC calculations and where $m_n$ is the nucleon mass.
The parameters of the first term, $a$ and $\alpha$, are mostly sensitive to the low density behavior of the EoS and are responsible of the two-nucleon part of the interaction.
The limits of $a$ and $\alpha$ are chosen so that we take into account all of the possible models from \cite{GCR12} (see Table \ref{tab:param_priors}).
On the other hand, the parameters of the second term, $b$ and $\beta$, are sensitive to the high density physics and control the three-nucleon interactions.
Furthermore, in our analysis we re-parameterized $b$ and $\beta$ in terms of $\mathcal{S}$ and $\mathcal{L}$.
Near the nuclear saturation density $n_0$ the symmetry energy of the neutron matter can be obtained from \eqref{eq:QMC_E} as
\be
\mathcal{S} \equiv \mathcal{S}(n_0) = E(n_0) - E_{\mathrm{nuc}}(n_0) = 16~\mathrm{MeV} + a + b,
\ee
where $E_{\mathrm{nuc}}(n_0) = -16~\mathrm{MeV}$ is the previously mentioned nuclear binding energy at the saturation density. 
For the pressure at the saturation density we obtain
\be
\mathcal{L} \equiv 3 n_0 \frac{\mathrm{d}\mathcal{S}(n)}{\mathrm{d}n} \Big|_{n=n_0} = 3(a\alpha + b\beta).
\ee
We also restrict the GCR model only up to nuclear saturation density as the validity of the model might not hold if a phase transition is present.

Above the saturation density $n_0$, a set of three piecewise polytropes are used and referred to as ``Model A'', similar to \cite{SLB13,SGF15}.
In this way, when parameterizing the high-density EoS we introduce three continuous power laws defining the pressure as
\be
P \propto \epsilon^{1+1/n},
\ee
as a function of the energy density $\epsilon$.
It has been shown that it is possible to model a wide range of theoretical model predictions with these kinds of simple polytropes with a typical rms error of about $4\%$ when compared to the actual numerical counterparts \citep{RLO09}.
In practice we can mimic theoretical models with up to three phase transitions because they will appear as successive polytropes with different indices.
Model A has 5 free parameters:
the first transition energy $\epsilon_1$ and the first polytrope index $n_1$, the second transition energy $\epsilon_2$ and the second polytrope index $n_2$, and a third polytrope index $n_3$ (see Table \ref{tab:param_priors} for the hard limits).
Additionally we, of course, require that $\epsilon_2 > \epsilon_1$ in order to avoid double-counting of the parameter space.
We have only 5 parameters (in contrast to 6) because the transition to the first polytrope is already fixed by the EoS at the saturation density.

An alternative for the high density EoS is a piecewise linear model referred to as ``Model C'' by \cite{SLB13,SGF15}.
Here the low density EoS is used up to $200~\mathrm{MeV}~\mathrm{fm}^{-3}$, after which four line segments are considered in the $\epsilon$ vs. $P$ plane at fixed energy densities of $400$, $600$, $1000$, and $1400~\mathrm{MeV}~\mathrm{fm}^{-3}$.
The linear relation between the two last regimes is extrapolated to higher densities, if needed.
  The Model C has four free parameters: $\Delta P_1\!=\!P(\epsilon\!=\!400) - P(\epsilon\!=\!200)$, $\Delta P_2\!=\!P(\epsilon\!=\!600) - P(\epsilon\!=\!400)$, $\Delta P_3\!=\!P(\epsilon\!=\!1000) - P(\epsilon\!=\!600)$, and $\Delta P_4\!=\!P(\epsilon\!=\!1400) - P(\epsilon\!=\!1000)$.
These pressure changes are always relative to the value of the pressure at the previous fixed point, effectively showing how sharply the pressure changes as a function of the energy density.
This second alternative model tends to favor strong phase transitions in the core so it is interesting to study the resulting differences between it and the more smoothly evolving polytropic model.

In total our EoS models have 9 (QMC + Model A) or 8 (QMC + Model C) free parameters.
In addition to the limits set to the priors (summarized in Table \ref{tab:param_priors}) some combinations of the parameters are rejected on a physical basis.
More precisely, we ensure that all
\begin{enumerate}
 \item mass-radius curves produce $2\Msun$ NS in line with the recent pulsar measurements \citep{AFW13, DPR10},
 \item obtained EoS are causal, i.e. $\mathrm{d}P/\mathrm{d}\epsilon > 0$ and
 \item EoSs are hydrodynamically stable everywhere, i.e. they have an
   increasing pressure with increasing energy density.
\end{enumerate}
In addition, during the computations, if any of the three masses obtained is larger than the maximum mass for the selected EoS, that realization is discarded and a new one is generated.

\subsection{EoS parameter results from the Bayesian analysis}

\begin{figure*}
\centering
\includegraphics[width=13cm]{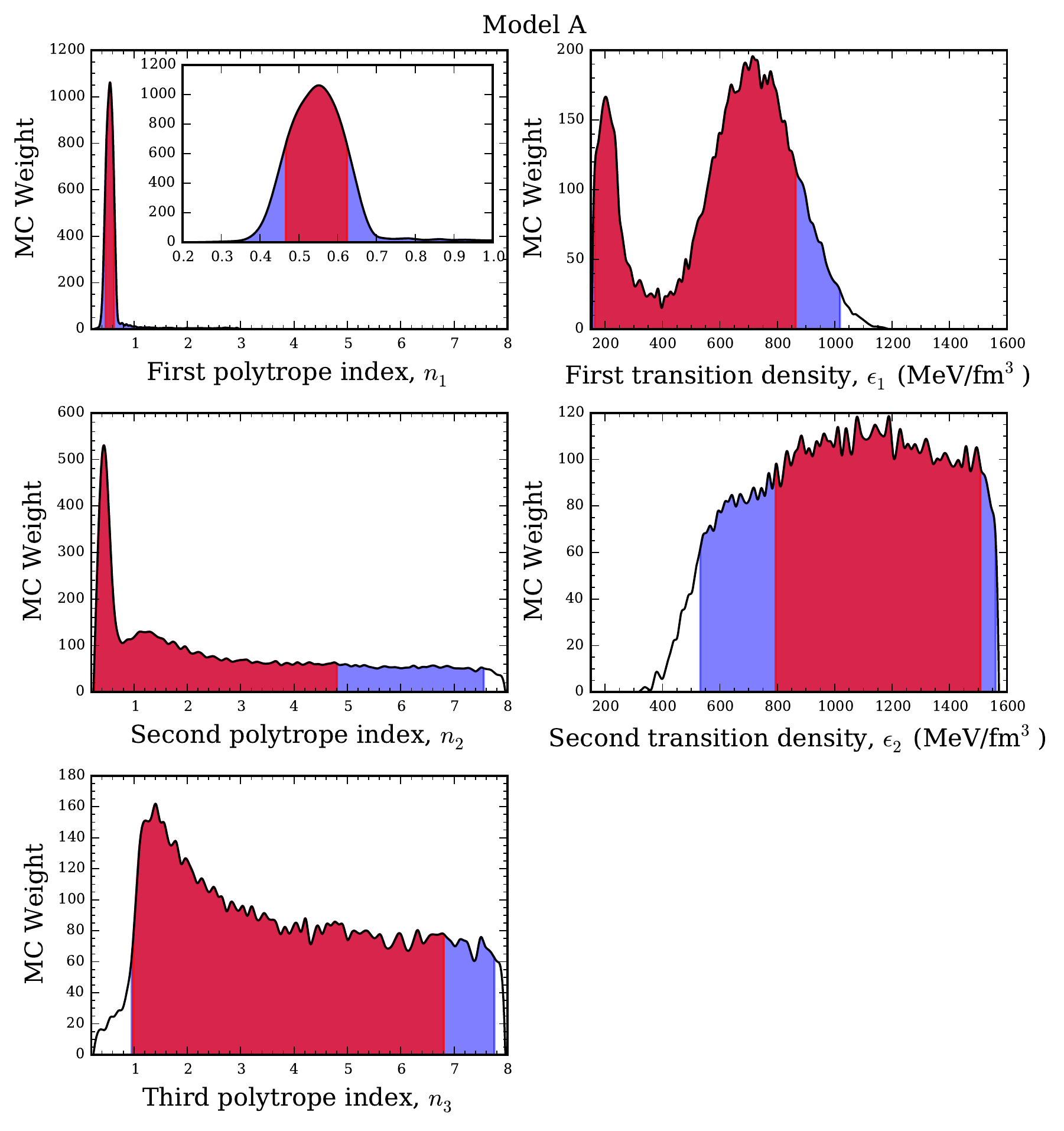}
\includegraphics[width=14.35cm]{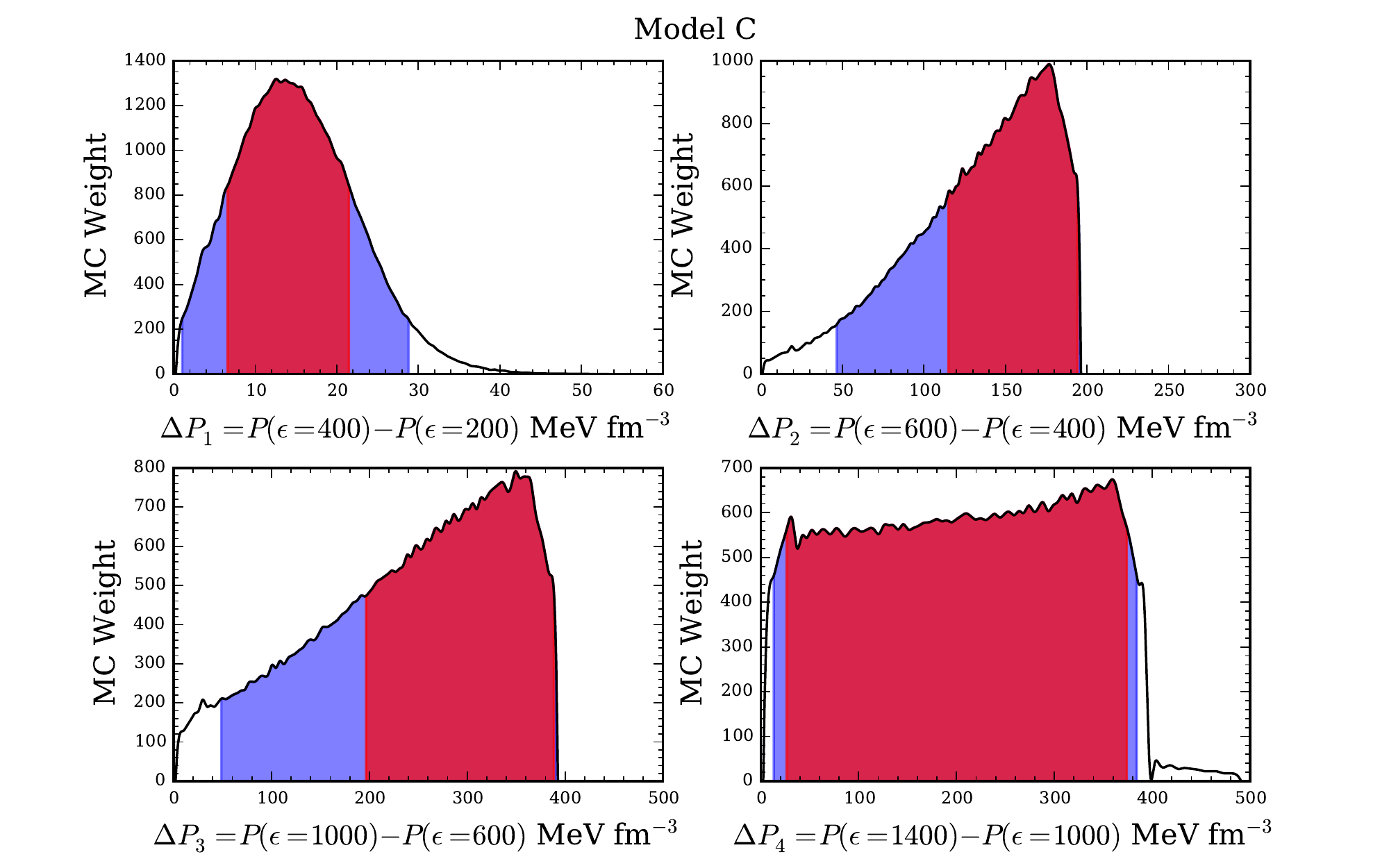}
\caption{\label{fig:EoSA}
  Histograms for the posterior distributions of the high-density Model A and Model C EoS parameters as implied by the data.
  In the first panel, the inset shows a magnified view of the histogram near $n_1 \approx 0.5$.
  Red shading corresponds to the $68\%$ and blue to the $95\%$ confidence regions of the parameters.
  Limits of the figures correspond to the hard limits set to the parameter prior distributions (see Table \ref{tab:param_priors}).
}
\end{figure*}

\begin{figure*}[ht]
\centering
\end{figure*}

The most probable value and the corresponding $1\sigma$ and $2\sigma$ limits for the EoSs are summarized in Table \ref{tab:param_posteriors} \citep[for the computation of the confidence regions, see][]{LS14b}.
We find that the posterior distributions for $a$ and $\alpha$, corresponding to the low-density EoS behavior (which is dominated by two-body interactions), are almost flat.
Thus, the neutron star observations do not constrain these parameters, as found previously~\citep{Steiner12cn}.
We find that the derivative of the nuclear symmetry energy $\mathcal{L}$ is only weakly constrained.
However, we do find a somewhat stronger constraint on the symmetry $\mathcal{S}$ than what has been obtained previously~\citep{SLB10}. The origin of this constraint is the combination of the neutron star data with the correlation between $\mathcal{S}$ and $\mathcal{L}$ found in quantum Monte Carlo results~\citep{GCR12}.
It is also remarkable that with both high-density models, Model A and Model C, the symmetry energy is constrained around $\mathcal{S} \approx 32~\mathrm{MeV}$, that is in good agreement with earthly measurements \citep[$\mathcal{S}=28-34~\mathrm{MeV}$,][]{KRB09}.
We, however, note that the parameters obtained here are to be considered as ``local'' quantities, as they are properties of the EoS only at densities close to the saturation density.

Histograms for the posterior distributions of the high-density parameters of the Model A are presented in Fig.~\ref{fig:EoSA}.
The index of the first polytrope $n_1$ is sharply peaking around $0.5$ corresponding to a polytropic exponent $\gamma_1 = 1 + 1/n_1 \sim 3$.
The large value implies a rather stiff EoS at supranuclear densities.
This first polytrope, corresponding to the $n_1$ index, is seen to continue all the way up to the first transition density at $\epsilon_1 \approx 700\pm150~\mathrm{MeV}\,\mathrm{fm}^{-3}$, i.e. around $4$ times the saturation energy density $\epsilon_0$.
In some realizations the transition is occuring already at around $\epsilon_1 \approx 200~\mathrm{MeV}\,\mathrm{fm}^{-3}$ but these cases correspond to the first sharp peak seen at $n_2 \approx 0.5$, i.e. we practically have only 2 polytropes spanning our energy density range. 
In this case, the role of the first polytrope is superseded by the second segment corresponding to index $n_2$.
In the opposite case, where all three polytropes span the grid the second polytrope index has values around $1.5$ (polytropic exponent $\gamma_2 \sim 1.7$) with a long tail extending all the way up to around $8.0$ (i.e. to the upper-limit of our prior).
What this means is that the data can not constrain the high-density behavior of the EoS very well.
As $n_2 > n_1$, it, however, indicates that some softening of the EoS is present at higher densities.
The third polytropic index $n_3$ is only weakly constrained to be $\gtrsim1$ (peaking around $1.5$) indicating either no phase transitions at all or additional softening as $n_3 > n_2 > n_1$.

Alternatively to the rather smoothly behaving polytrope model, we take the piecewise linear Model C that can shows strong phase transitions.
Instead of extending the $n_2$ and $n_3$ parameter upper-limits of Model A, that would create an apparent bias for softer EoS, we can characterize the effect of softer EoSs by applying this piecewise parameterization to the data, too.
The histograms of the posterior distributions of the obtained pressure differences (at the fixed transition densities $\epsilon = 400$, $600$, $1000$, and $1400~\mathrm{MeV}\,\mathrm{fm}^{-3}$) are presented in Fig. \ref{fig:EoSA}.
For the first segment from $\epsilon_{\mathrm{trans}} = 200$ to $\epsilon_1 = 400~\mathrm{MeV}$ the difference is tightly constrained around $\Delta P_1 = 15~\mathrm{MeV}\,\mathrm{fm}^{-3}$ with an almost symmetric Gaussian distribution with tails of $1\sigma \approx 7 ~\mathrm{MeV}\,\mathrm{fm}^{-3}$.
This introduces a strong phase transition to the EoS as the pressure changes by only  a little.
After this possible transition the EoS hardens out and for $\Delta P_2$ and $\Delta P_3$ (at $600$ and $1000~\mathrm{MeV}\,\mathrm{fm}^{-3}$) the posterior distributions are considerably asymmetric toward larger pressure changes peaking at $180$ and $350~\mathrm{MeV}\,\mathrm{fm}^{-3}$, respectively. 
For the final possible line segment, the EoS appears almost unconstrained.
The sharp cutoff at higher pressures for $\Delta P_2$, $\Delta P_3$, and $\Delta P_4$ appears because we rule out EoSs where the speed of sound exceeds the value of the speed of light.
The modest high-value tail for $\Delta P_4$ originates from a small group of EoSs where the central value does not exceed $1000~\mathrm{MeV}\,\mathrm{fm}^{-3}$ and hence every value of the parameter is allowed.

\subsection{Predicted EoS}

\begin{figure*}
\centering
\includegraphics[width=15cm]{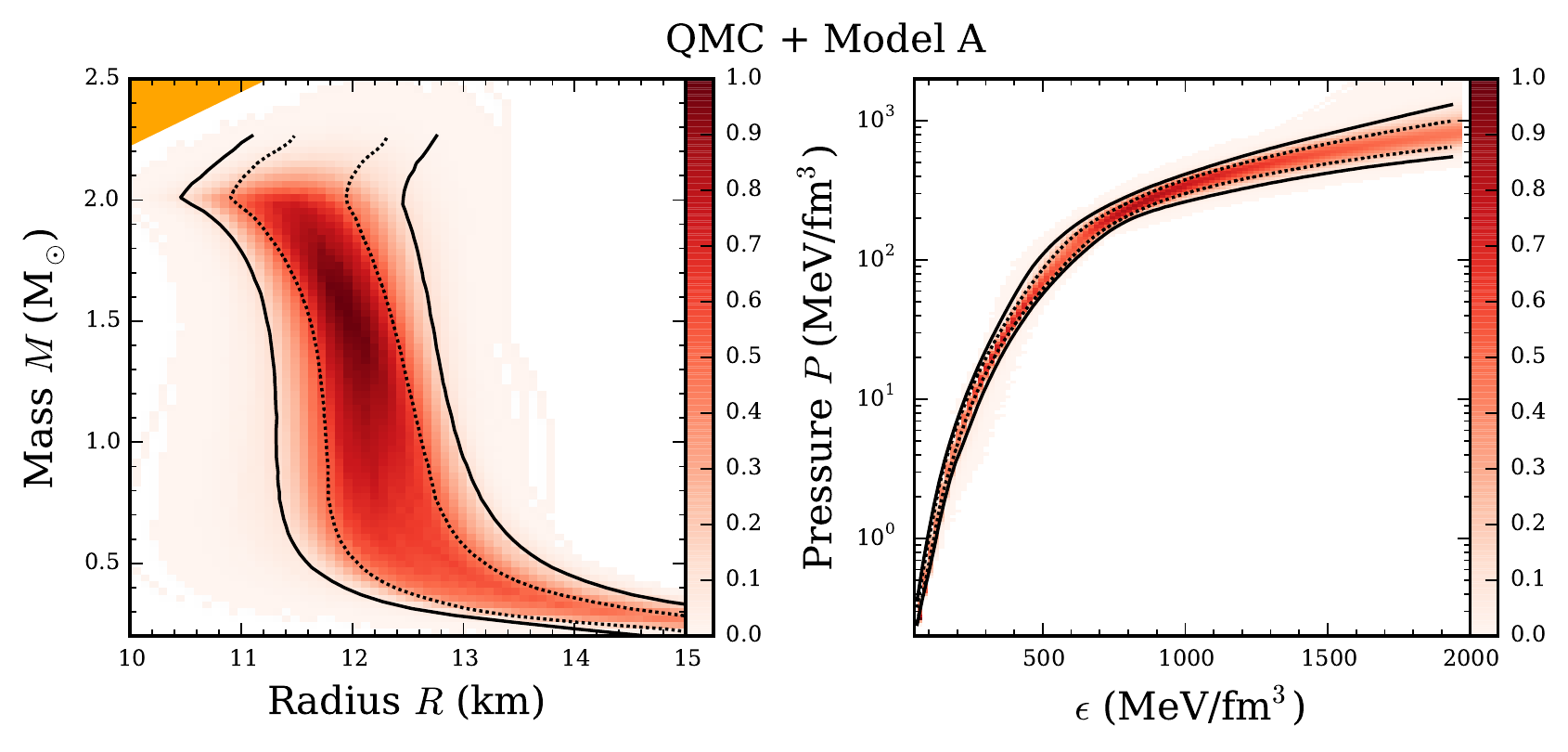}
\includegraphics[width=15cm]{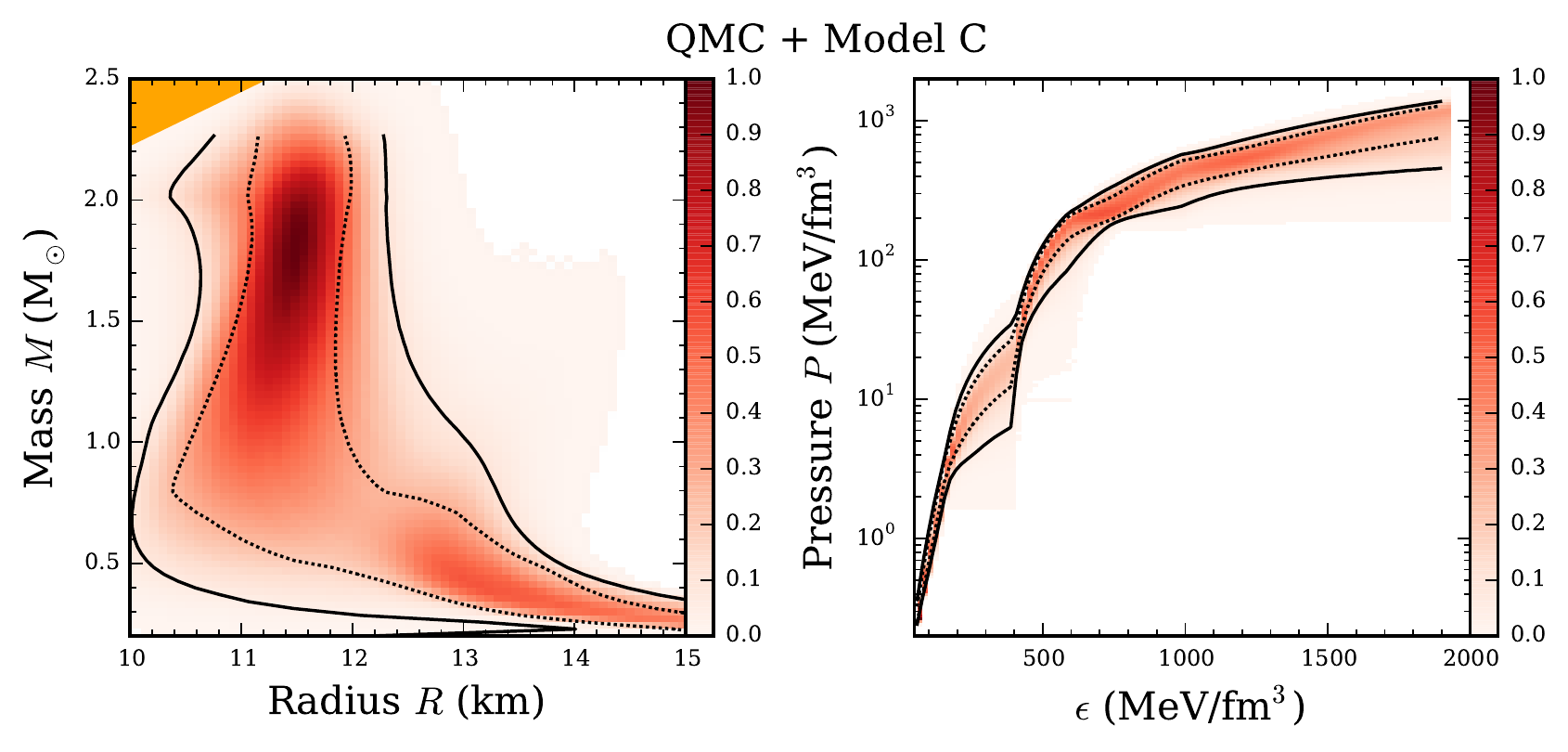}
\caption{\label{fig:MR_PE}
  Obtained EoS constraints in the $M-R$ (left panel) and in the $P-\epsilon$ plane (right panel).
  Upper panels correspond to the QMC + Model A and bottom panels to the QMC + Model C EoS.
  Red color indicates the probability density and black lines show the $68\%$ (dotted) and $95\%$ (solid) confidence limit contours.
}
\end{figure*}

The predicted EoS obtained from the X-ray burst data is shown in Fig.~\ref{fig:MR_PE} in $M-R$ and $P-\epsilon$ planes.
Each panel in the figure displays an ensemble of 1d-histograms over a fixed grid in one of the axes (note that this is not quite the same as a 2d-histogram).
The right panels present the ensemble of histograms of the pressure for each energy density.
This was computed in the following way:
for each energy density, we determined the histogram bins of pressure which enclose $68\%$ and $95\%$ of the total MC weight.
The location of those regions for each 1d-histogram are outlined by dotted and solid curves, respectively, and these form the contour lines.
These $1\sigma$ and $2\sigma$ contour lines give constraints on the pressure as a function of the energy density as implied by the three NS data sets (see also Table \ref{tab:e_P_A} and \ref{tab:e_P_C}).
Very high density behavior between the Model A and C are seen to be similar as both of the models appear rather soft in this regime.
At these very high energy density regions it is actually the maximum mass requirement that constraints the pressure evolution \citep[see][for more extensive discussion]{SLB13}.
Most striking difference occur at lower energy densities where the sharp phase transition in the QMC $+$ Model C EoS in seen to produce large scatter to the pressure at supranuclear densities (around $\epsilon \approx 400~\mathrm{MeV}\,\mathrm{fm}^{-3}$).

Similarly, the left panels of Fig.~\ref{fig:MR_PE} presents our results for the predicted mass-radius relations.
These panels present the ensemble of histograms of the radius over a fixed grid in neutron star mass with $1\sigma$ and $2\sigma$ constraints presented with dotted and solid lines, respectively, similar to the right panels.
See also Tables \ref{tab:m_r_A} and \ref{tab:m_r_C} that summarize these contour lines as well as give the most probable $M-R$ curve.
The width of the contours at masses higher than about $1.8~\Msun$ tends to be large because the available NS mass and radius data in our sample generally imply smaller masses which in turn leads to weaker constraints.
The obtained EoS for the QMC $+$ Model A has a predicted radius that is almost constant over the whole range of viable masses.
The radius is constrained between $11.3 - 12.8~\mathrm{km}$ for $M=1.4~\Msun$ ($2\sigma$ confidence limits).
Constraints this strong are obtained because the combination of weak (or non-existent) phase transitions and the NS mass and radius measurements from the cooling tail method compliment each other well:
Cooling tail measurements are elongated along the constant Eddington temperature curve that stretches from small mass and small radius to large mass and large radius.
On the other hand, the assumption of weak phase transitions in the EoS forces the radius to be almost independent of the mass.
This assumption of constant radius then eliminates some of the uncertainties present in the cooling tail measurements (mostly due to the unknown distance) as each individual measurement is required to have (almost) the same radius. 

With the QMC $+$ Model C, on the contrary, the first phase transition at supranuclear densities produce slightly skewed mass-radius curve to compensate the cooling tail burst data that is elongated along the constant Eddington temperature.
With this possible phase transition present in the EoS the mass-radius curve is then able to support high-mass NSs with radius of about $R \approx 11.6~\mathrm{km}$ and low-mass stars with smaller radii of around $R \approx 11.3~\mathrm{km}$ simultaneously.
The phase transition also causes a large scatter to the radius below $1~\Msun$ as the exact location of the turning point where the radius starts to increase again, cannot be constrained from the available data.
Because of these uncertainties originating from the possible phase transition, the Model C shows a much larger scatter in the predicted radii at small masses.
The radius is constrained between $10.5 - 12.5~\mathrm{km}$ for $M=1.4~\Msun$ ($2\sigma$ confidence limits).

\subsection{Individual mass and radius results for the NSs}

\begin{table}
\caption{Most probable values for masses and radii for NSs constrained to lie on one mass versus radius curve}
\centering
\begin{small}
\begin{tabular}[c]{l c c}
 \hline
 \hline
         & $M$          & $R$\\
  Source & ($\Msun$)  & (km)\\[5pt]
  \hline
  \multicolumn{3}{c}{QMC  $+$ Model A}\\
4U 1702$-$429        & $1.8^{+0.2~(+0.3)}_{-0.3~(-0.6)}$ & $11.9^{+0.4~(+0.8)}_{-0.6~(-1.1)}$ \\[5pt]
4U 1724$-$307        & $1.5^{+0.4~(+0.6)}_{-0.3~(-0.4)}$ & $12.0^{+0.5~(+0.8)}_{-0.5~(-1.0)}$\\[5pt]
SAX J1810.8$-$2609   & $1.4^{+0.4~(+0.6)}_{-0.4~(-0.5)}$ & $12.0^{+0.4~(+0.8)}_{-0.5~(-1.0)}$\\[5pt]
\hline
\multicolumn{3}{c}{QMC  $+$ Model C}\\
4U 1702$-$429        & $1.8^{+0.2~(+0.3)}_{-0.3~(-0.7)}$ & $11.4^{+0.5~(+1.0)}_{-0.5~(-1.0)}$ \\[5pt]
4U 1724$-$307        & $1.3^{+0.6~(+0.7)}_{-0.3~(-0.5)}$ & $11.4^{+0.6~(+1.0)}_{-0.5~(-1.1)}$ \\[5pt]
SAX J1810.8$-$2609   & $1.2^{+0.6~(+0.9)}_{-0.3~(-0.4)}$ & $11.5^{+0.5~(+1.0)}_{-0.6~(-1.2)}$ \\[5pt]
\hline
\end{tabular}
\end{small}
\label{tab:mr_sep}
\begin{center}
  {\small{
      Note: Errors correspond to the $68\%$ and $95\%$ (in parenthesis) confidence levels.\\
}}
   \end{center}
\end{table}

\begin{figure*}[!p]
\centering
\includegraphics[width=20cm]{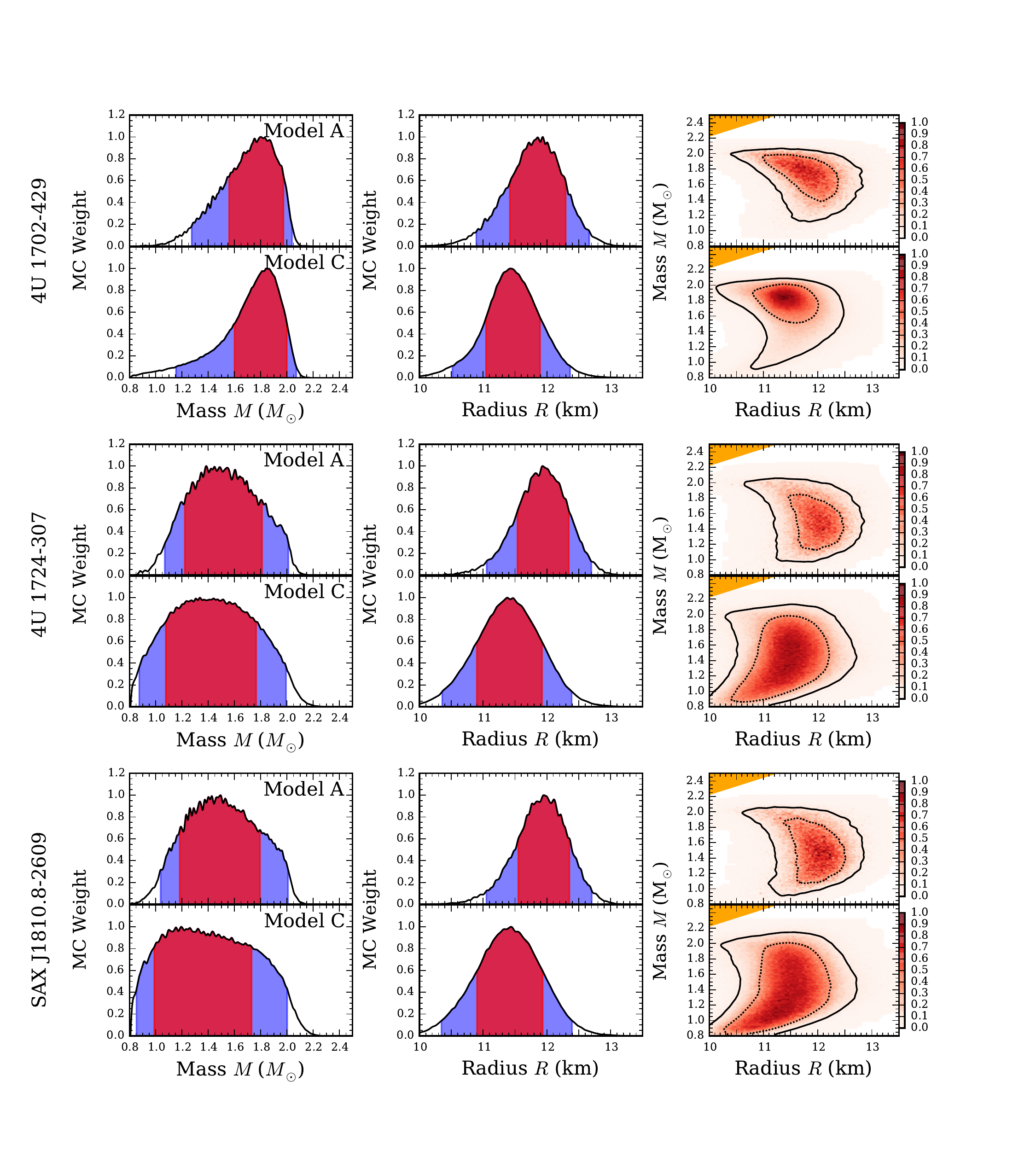}
\caption{\label{fig:mr_sep}
  Individual mass and radius constraints for the three neutron stars used in the analysis.
  Left panel shows the projected mass and the middle panel the projected radius histograms.
  Red shading corresponds to the $68\%$ and blue to the $95\%$ confidence regions of these parameters.
  Right panels shows the full 2-dimensional mass and radius probability distributions.
  Contours of $68\%$ (dotted, black line) and $95\%$ (solid, black line) confidence regions are also shown.
}
\end{figure*}

The combination of several neutron star mass and radius measurements with the assumption that all neutron stars must lie on the same mass-radius curve puts also a significant constraints to the mass and radius of each object.
The resulting $M$ and $R$ constraints for each object are given in Fig.~\ref{fig:mr_sep} and summarized in Table~\ref{tab:mr_sep}.
In general, the QMC $+$ Model A EoS tends to favor slightly larger masses and larger radii than compared to the QMC $+$ Model C.
With the polytropic Model A, the resulting mass is tightly constrained around $M \approx 1.5~\Msun$ for 4U 1724$-$307 and for SAX J1810.8$-$2609.
Slightly larger mass of around $M \approx 1.8~\Msun$ is obtained for the remaining 4U 1702$-$429.
Resulting radii are constrained to be around $R \approx 12.0~\mathrm{km}$ for each source.
With the piecewise linear Model C the mass is about $M \approx 1.3~\Msun$ for the 4U 1724$-$307 and SAX J1810.8$-$2609 and, again, around $M \approx 1.8~\Msun$ for 4U 1702$-$429.
The obtained radii are located around $R \approx 11.4~\mathrm{km}$.
Because of the uncertainties from the possible phase transition occuring in the Model C, the resulting mass and radius constraints for each source are also much more loose.

\section{Discussion}\label{sect:disc}

In this paper we have used the cooling tail method to constrain the mass and radius of three NS X-ray bursters: 4U 1702$-$429, 4U 1724$-$307, and SAX J1810.8$-$2609.
Special care was taken to use only the passively cooling bursts as theoretical calculations of the color-correction factor $f_{\mathrm{c}}$ affecting the emerging spectra do not take any external heating into account.
In practice this means that the blackbody normalization $K$ is required to evolve during the burst (because $K^{-1/4} \propto f_{\mathrm{c}}$; see, e.g., \citealt{PNK14, KNL14}) and is indeed what we observed for the bursts in our sample.

First we introduced a new Bayesian cooling tail method and assumed uniform $M$ vs. $R$ priors in our analysis (instead of uniform $F_{\mathrm{Edd}}$ and $A$ priors).
By marginalizing over the $M$, $R$, and $X$ prior distributions we also got distance estimates for our sources.
One advantage here is that measurements like this are basically done in the X-ray band where interstellar extinction does not play such an important role, hence reducing possible model dependencies originating e.g. from selection of the interstellar extinction model.
One should, however, note that in our case completely different kinds of model dependencies are present, related, for example, to the uncertain composition of the accreted material (i.e. the value of the hydrogen mass fraction) or to the X-ray burst selection.
Unfortunately, only 4U 1724$-$307 has some distance measurements that we can compare against, as it is located in the globular cluster Terzan 2.
Distance estimates to this source range from $5.3$ to $7.7~\mathrm{kpc}$ \citep[][using either extinction model valid for red stars or an average value from some large sample, respectively]{OBB97} in addition to the more recent measurement of $7.4\pm0.5~\mathrm{kpc}$ using near-IR observations of red giant branch stars \citep{VFO10}.
We note that our distance constraints are consistent with the lower-limit end of the measurements as $D_{\mathrm{max}} \approx 6~\mathrm{kpc}$.
This value is, however, dependent on our selection of hydrogen mass fraction and should not be interpreted as a strict limit.
If we would decrease the hydrogen mass fraction (and hence increase the $D_{\mathrm{max}}$ value) our resulting radii would also rapidly increase, creating tension between our other measurements.
Interestingly, if we would impose a cut around $5.3~\mathrm{kpc}$ into our distance prior, our $M$ vs. $R$ results would be tightly constrained around $R = 12.0\pm0.3~\mathrm{km}$ and $M = 1.5 \pm 0.2~\Msun$ as the new distance prior would remove the low-mass solutions.
For 4U 1702$-$429 and SAX J1810.8$-$2609 we constrained the distance to be around $5.6_{-0.9}^{+0.6}~\mathrm{kpc}$ and $4.3_{-1.0}^{+0.6}~\mathrm{kpc}$ ($2\sigma$ confidence limits), respectively.

Mass and radius constraints of 4U 1724$-$307 and SAX J1810.8$-$2609 are found to be almost identical with the radius confined between about $11 - 13$ km.
Both of these sources are also best modeled by a solar-like composition with almost zero metallicity (SolA001 model).
The best-fit model for the third, remaining source 4U 1702$-$429 consists of pure helium.
This implies a hydrogen-poor companion like a white dwarf, that in turn, implies a compact binary system in order for the accretion to proceed via Roche lobe overflow.
Best-fit values for the radius of this source (with $X=0$) give $R \approx 13~\mathrm{km}$ at around $M = 1.5~\Msun$, a slightly larger value compared to the two other sources.


Some physical uncertainties are also still present in the X-ray burst $M-R$ measurements.
For example, no rotation is taken into account in the current work.
Rotation affects the emerging spectrum owing to the various special relativistic effects \citep[see, e.g.,][]{BOP15} but most importantly because the radius of the star increases at the equator and decreases at the poles as the star becomes oblate \citep{MLC07, BBP13}
It, however, also introduces two new unknown free parameters to the fits: the spin period and the inclination.
In many cases these parameters are not known a priori (especially the inclination).
In addition, the flux distribution over the surface of the NS is unknown.
The rotation does not, however, have a considerable effect for the $M$ and $R$ constraints when the spin frequency is moderate ($\lesssim 400~\mathrm{Hz}$).
Luckily, most of the X-ray bursters detected seem to have a frequency around this range.
In our case, only 4U 1702$-$429 has a measured spin period of $329~\mathrm{Hz}$, i.e. it is a slow rotator and no major corrections are expected.
For the two other sources, no burst oscillations nor persistent millisecond pulsations were detected and so the spin period is unknown.
Other uncertainties may arise from the composition that can have an impact on the color-correction factors \citep{NSK15}.
Stratification may create an almost pure hydrogen layer or, in the opposite situation, strong convection might mix up the whole photosphere thoroughly, including the burning ashes consisting of heavier elements \citep[see e.g.,][]{WBS06, Malone11, Malone14}.
There are, however, convincing implications that these do not affect our current measurements as the observations of the normalization $K$ do really seem to follow the theoretical atmosphere models with the simple solar-like (or pure He) compositions very closely.
Additional confirmation is also obtained from the corrected temperatures that follow the $F \propto T_{\mathrm{eff}}^4$ law implying that the values of $f_{\mathrm{c}}$ used are correct. 
In the end, the distance is still our biggest source of uncertainty in the measurements.

Similarly to the physical uncertainties, some technical sources of systematic error are present.
For example data selection for individual bursts plays a role:
fitting bursts from the touchdown down to only half of the touchdown flux (instead of the $20\%$ used here) increases the uncertainties, as one would expect because we use less data, but also increases the radius as much as about $800~\mathrm{m}$.
  Also by refining the cooling tail method (where $f_{\mathrm{c}} - \ell$ is used in the fitting procedures) into a more accurate two parameter treatment (where our model is $w^{-1/4} - (w f_{\mathrm{c}}^4) \ell$ and $w$ is the dilution factor that was previously assumed to be $w \approx f_{\mathrm{c}}^{-4}$) the radius is seen to increase by about $500~\mathrm{m}$ (Suleimanov et al., in prep).
  Because both of these effects act to increase our radius we can understand our current measurements as a lower limit to the real radius.

The results presented here constitute the first observational NS $M-R$ constraints for 4U 1702$-$29 and SAX J1810.8$-$2609 using \textit{RXTE}/PCA data.
In addition, we constrained the compactness for the NS in 4U 1724$-$307 that has already been previously analyzed by \citet{SPRW11} and \citet{OPG15}.
These three measurements were then used to create our parameterized EoS of cold dense matter.

In general, our EoS $M-R$ constraints are slightly different when compared to \citet{OPG15} results as our radii tends to be somewhat larger in between about $10.6 - 12.4~\mathrm{km}$ for Model C and $11.2 - 12.7~\mathrm{km}$ for Model A as compared to their measurement of $10.1 - 11.1~\mathrm{km}$ for $M = 1.5~\Msun$.
Note also that the parameterization of the EoS in \citet{OPG15} is closer to our QMC + Model A polytropic formalism.
One possible cause for the difference might be the data selection:
the PRE burst we analyzed in this paper occurred in the hard spectral state with a low persistent flux level, whereas most of the bursts analyzed in \citet{OPG15} occur in the soft spectral state and at higher persistent flux.
As mentioned in the introduction, hard state X-ray bursts - such as the ones analyzed in this paper – tend to follow the NS atmosphere model predictions, whereas the soft state bursts never follow them \citep[see Fig. \ref{fig:cooling_tails} and also][ for discussion]{SPRW11, PNK14,KNL14}.
We therefore argue that in the soft state bursts there is an additional physical process (i.e. the spreading boundary layer) acting on the burst emission, that causes the assumptions on the visibility of the entire NS to break down and the value of the color-correction factor to be different from what is predicted by the passively cooling atmosphere models.
One should also notice the completely different shape of the $M-R$ contours obtained in \citet{OPG15} where all of the $M-R$ points are close to the region where only one mass-radius solutions exists \citep[see, e.g.,][for more in depth discussion about this]{PNK14}, indicating that the lower-limit for the distance is close to the maximum distance $D_{\mathrm{max}}$.
This also creates tension for the solutions to exists on higher masses as the one-solution point lies on the $R = 4GM/c^2$ line \citep[see, e.g.,][]{OP15}.
In our case, no such a tension exists, that in turn leads in to much bigger $M-R$ errors.
  
  Specifically for the 4U 1724$-$307 we obtain $R \approx 12.0\pm0.5~\mathrm{km}$ whereas \citet{OPG15} obtains $\sim 12.2~\mathrm{km}$.
  Here the difference is already well inside error-limits. 
\citet{SPRW11,SPW11} have also analyzed 4U 1724$-$307 using a different hard-state burst than what is in our sample and the resulting radii are considerably different ($R > 14~\mathrm{km}$).
There is, however, a possibility that the atmosphere during this burst might not consist of hydrogen and helium only, but is enriched with the nuclear burning ashes.
These ashes would then affect the measured color-correction factor considerably \citep[see][and Appendix \ref{sect:1724old} for more in-depth discussion about this burst]{NSK15}.
In the recent analysis of 4U 1608$-$52 by \citet{PNK14} they also used only the hard-state bursts from the source to constrain the mass and the radius.
The resulting radii were, again, somewhat larger at around $R \gtrsim 13~\mathrm{km}$ for $M = 1.4~\Msun$.
We, however, note that 4U 1608$-$52 is a fast rotator with an observed spin frequency of $620$ Hz \citep[largest observed for an X-ray burster;][]{MCG02}.
This means that for an exact and reliable $M-R$ analysis, a rotation-modified cooling tail method should be used (Suleimanov et al., in prep%
\footnote{Preliminary results with rotating NSs and rotation-modified cooling tail method show a reduction of the measured radius by about $1-2~\mathrm{km}$ for fast rotators mostly due to the oblate emitting shape of the star and the varying flux distribution on the surface.}; %
 see also \citealt{BOP15}).

In the second part of the paper, we combined our separate cooling tail measurements to obtain a parameterized EoS of the cold dense matter inside NSs.
By utilizing a Bayesian framework, we studied the constraints that are possible to set to the EoS parameters such as the symmetry energy $\mathcal{S}$ and its density derivative $\mathcal{L}$.
We also relaxed our EoS priors by studying two different high-density models: Model A consisting of polytropes and Model C with piecewise linear segments in the $\epsilon - P$ plane.
Here the Model C with linear segments supports strong phase transitions in the supranuclear densities and indeed our resulting EoS is seen to have large transition around $\epsilon \sim 400~\mathrm{MeV}\,\mathrm{fm}^{-3}$ to support both large radius for large-mass and smaller radius for small-mass stars.
Model A, on the other hand, gives much tighter constraints as the pressure in the core evolves rather smoothly and hence the resulting radius is almost independent of the mass.
This restriction of having a constant radius then asserts tight limits to the obtained $M$ vs. $R$ results.

Nevertheless, all of the obtained constraints seem to favor an EoS that produces a radius in between $10.5-12.8~\mathrm{km}$ ($95\%$ confidence limits) for a mass of $1.4~\Msun$.
One should, however, keep in mind that in reality the aforementioned systematic errors might increase these limits slightly.
It is also interesting to note that we do not necessarily need a far better precision for the $M-R$ measurements as interesting constraints can already be obtained from the existing observations.
Encouragingly, the astrophysical constraints also seem to agree with nuclear physics experiments, theoretical studies and heavy-ion experiments of neutron matter \citep{LL13}.
In addition to the EoS constraints of the super-dense matter we also get additional constraints for individual mass and radius measurements of the single NSs.
By assuming that all of the sources must lie on the same $M$ vs. $R$ curve we can set stronger constraints than what is possible with only the cooling tail method alone.
Especially the results with the QMC $+$ Model A shows how the unknown distance uncertainties (i.e. elongated $M$ vs. $R$ contours along the constant Eddington temperature) can be reduced by combining different sources with this kind of joint EoS fit.
In the end, the mass, for example, can be constrained to an accuracy of about $0.2 - 0.3~\Msun$ ($1\sigma$ confidence level) from the original limits spanning from $M=0.8$ to $2.2~\Msun$.

Future prospects include an extended study of uninformative (for example, Jeffrey's) priors in $M-R$ space, both for the EoS and cooling tail fit parameters.
This should be done to ensure that we do not implicitly and unknowingly transfer information to the mass and radius posteriors that already have many uncertainties present due to the poor measurements.
Other obvious prospect is the addition of all possible $M-R$ measurements including, but not restricted to, other X-ray bursting sources, quiescent LMXBs, thermally emitting isolated NSs, neutron star seismology results, pulse profiles in X-ray pulsars, moment of inertia and crust thickness measurements.

\section*{Acknowledgments}
Authors would like to thank M.C.~Miller and J.~Lattimer for valuable discussions and comments. We also thank the anonymous referee for extremely thorough inspection of the manuscript and helpful comments that significantly improved the paper.
This research was supported by the V\"ais\"al\"a Foundation and by the University of Turku Graduate School in Physical and Chemical Sciences (JN).
VS was supported by the German Research Foundation (DFG) grant WE 1312/48-1.
JJEK acknowledges support from the ESA research fellowship programme.
JP acknowledges funding from the Academy of Finland grant 268740.
JN and JJEK acknowledge support from the Faculty of the European Space Astronomy Centre (ESAC).
This research was undertaken on Finnish Grid Infrastructure (FGI) resources.


\clearpage

\begin{appendix}

\section{Two hard-state burst from 4U 1724$-$307}\label{sect:1724old}

\begin{figure}[!htb]
\centering
\includegraphics[width=9.2cm]{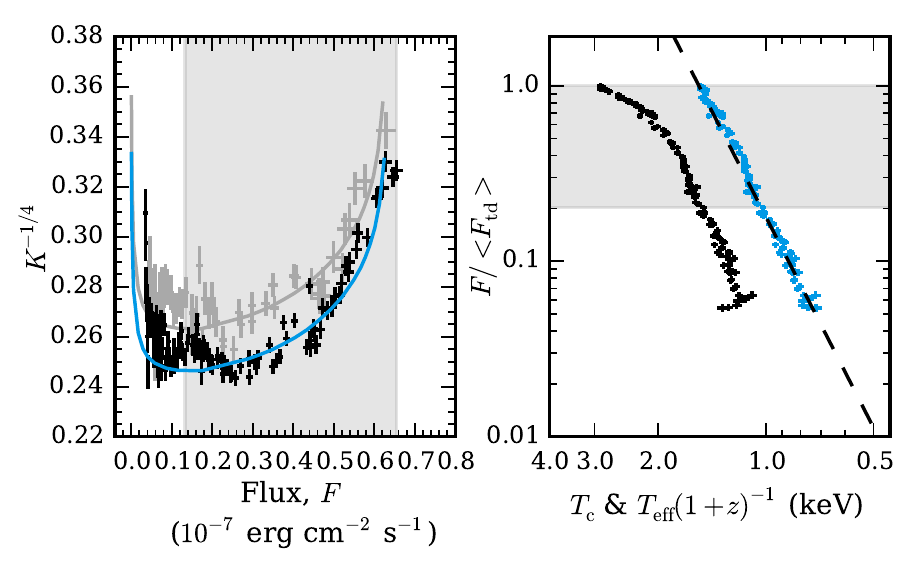}
\caption{\label{fig:1724old}
  \textit{Left panel:} Cooling tail in the $F \propto L/L_{\mathrm{Edd}}$ vs $K^{-1/4} \propto f_{\mathrm{c}}$ plane with 1$\sigma$ errors presented by black crosses for a hard-state PRE burst from 4U 1724$-$307 analyzed by \citet{SPW11}.
  Best-fit theoretical atmosphere model is shown with the blue curve (SolA001).
  A PRE burst from the same source (used in the this paper) is also presented with gray crosses and gray curve. 
  \textit{Right panel:} Temperature evolution of the \citet{SPW11} burst.
  Blackbody temperature $T_{\mathrm{bb}}$ is shown for the cooling tail with black crosses. 
  Blue crosses show the color-corrected temperatures $T_{\mathrm{eff}}(1+z)^{-1}$.
  The straight inclined lines show a powerlaw with an index of 4 corresponding to the $F \propto T^4$ relation.
}
\end{figure}

The cooling tail method has been previously applied to another hard-state burst (obsid: 10090-01-01-021) from 4U 1724$-$307 \citep{SPRW11, SPW11}.
In this case, a rather large radius in excess of $13$ km was obtained.
Since then, there has been a second hard-state burst from the same source that does not, however, follow the same track in the $K^{-1/4}$--$F$ plane (this new burst is the one we use in our analysis; see Fig. \ref{fig:1724old}).
We also note that the burst data presented in \cite{SPW11,SPW12} was reduced using an older {\it RXTE} data reduction pipeline without deadtime correction of the PCU detectors.
Since 2010, the \textit {RXTE}  team also introduced a correction to the effective area of the instrument that changed the measured flux by up to $10\%$.
With our current (and final) {\it RXTE}  data reduction methods (see Sect.~\ref{sect:data}) and with the new color-correction factors from \citet{SPW12} the older burst results in a radius of around $15$ km at $1.4~\Msun$ when the cooling tail method is applied with the solar composition model (SolA001).
With pure hydrogen composition the radius is brought down to the $\sim 14$ km range.
We also note that in \citet{SPW11,SPW12} the first five most luminous points near the touchdown are ignored in the fit.
Ignoring these points, however, leads to smaller Eddington flux and, hence, even larger radius of around $16-17~\mathrm{km}$ at $1.4~\Msun$ with our current data and new analysis methods.

The old burst also has poor $\chi^2$ values at the low luminosity tail (below about $< 0.5 F_{\mathrm{td}}$) originating from an additional powerlaw-like distribution of high-energy photons that cannot be described by the blackbody law.
This, in turn, might imply an additional heating of the surface by infalling material from the hot accretion flow.
The extra heating does not, however, explain the discrepancy between the two bursts at higher luminosities.
At these higher luminosities, it is possible that the NS atmosphere does not consist of hydrogen and helium only, but is enriched with the nuclear burning ashes \citep{WBS06}.
The presence of these ashes can then have a substantial impact on the color-correction factor reducing it by as much as $40\%$ \citep{NSK15}.
This is also supported by the exceptionally long duration of the burst ($\sim 100~\mathrm{s}$ compared to $\sim 30~\mathrm{s}$ for the other bursts in our sample) that might drive up the convection.
Furthermore, the temperature evolution of this burst also show some deviations from the $L \propto T^4$ law (see right panel of Fig.~\ref{fig:1724old}) that might imply inconsistencies with pure solar composition.
Because to these uncertainties, we choose to leave this burst out from our sample and only use the newer hard-state burst from the source that is consistent with passive cooling.

\clearpage
\section{Uniform $F_{\mathrm{Edd}}$ and $A$ priors}\label{sect:AppendixB}

Instead of assuming uniform $M$ vs. $R$ priors we can assume uniform $F_{\mathrm{Edd}}$ and $A$ priors (and compute $M$ and $R$) as was done in our previous work \citep[see, e.g.,][]{PNK14}.
This kind of selection of priors, however, turns out to be problematic in many ways.

First of all, we cannot truly introduce $\log g$ dependency into to the models as we do not know $M$ nor $R$ beforehand.
In this case, we are left to iteratively selecting the best-fit surface gravity from the three basic values ($\log g = 14.0$, $14.3$, and $14.6$) instead of interpolating between the models, as is done in the current work.

\begin{figure}[!htb]
\centering
\includegraphics[width=9cm]{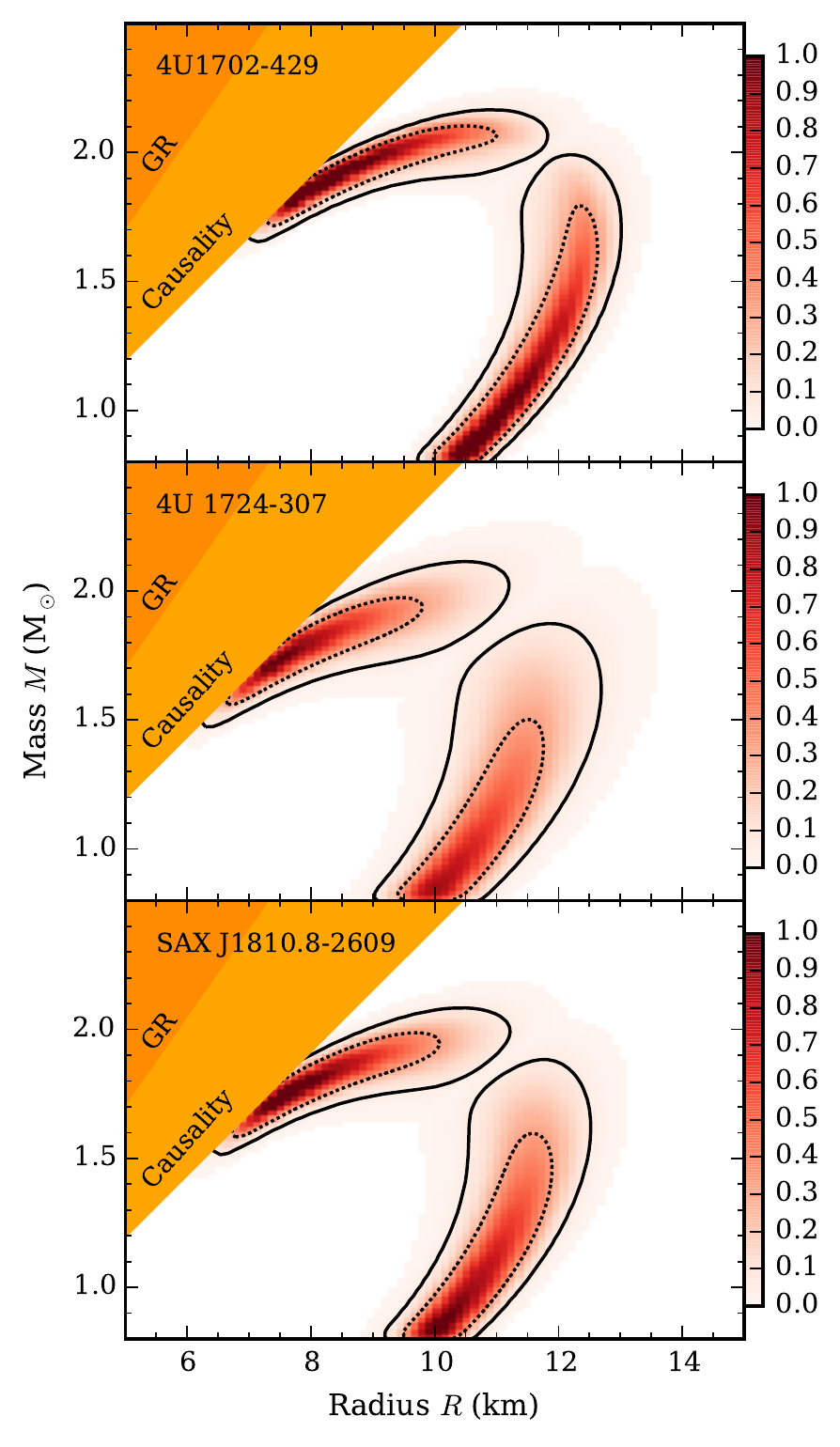}
\caption{\label{fig:MR_Appendix}
  Mass-radius constraints for the sources from the hard state PRE bursts assuming uniform $F_{\mathrm{Edd}}$ and $A$ priors.
Constraints are shown by $68\%$ (dotted line) and $95\%$ (solid line) confidence level contours.
The upper-left region is excluded by constraints from the causality and general relativistic requirements \citep{HPY07,LP07}. 
}
\end{figure}

\begin{figure*}
\centering
\includegraphics[width=19cm]{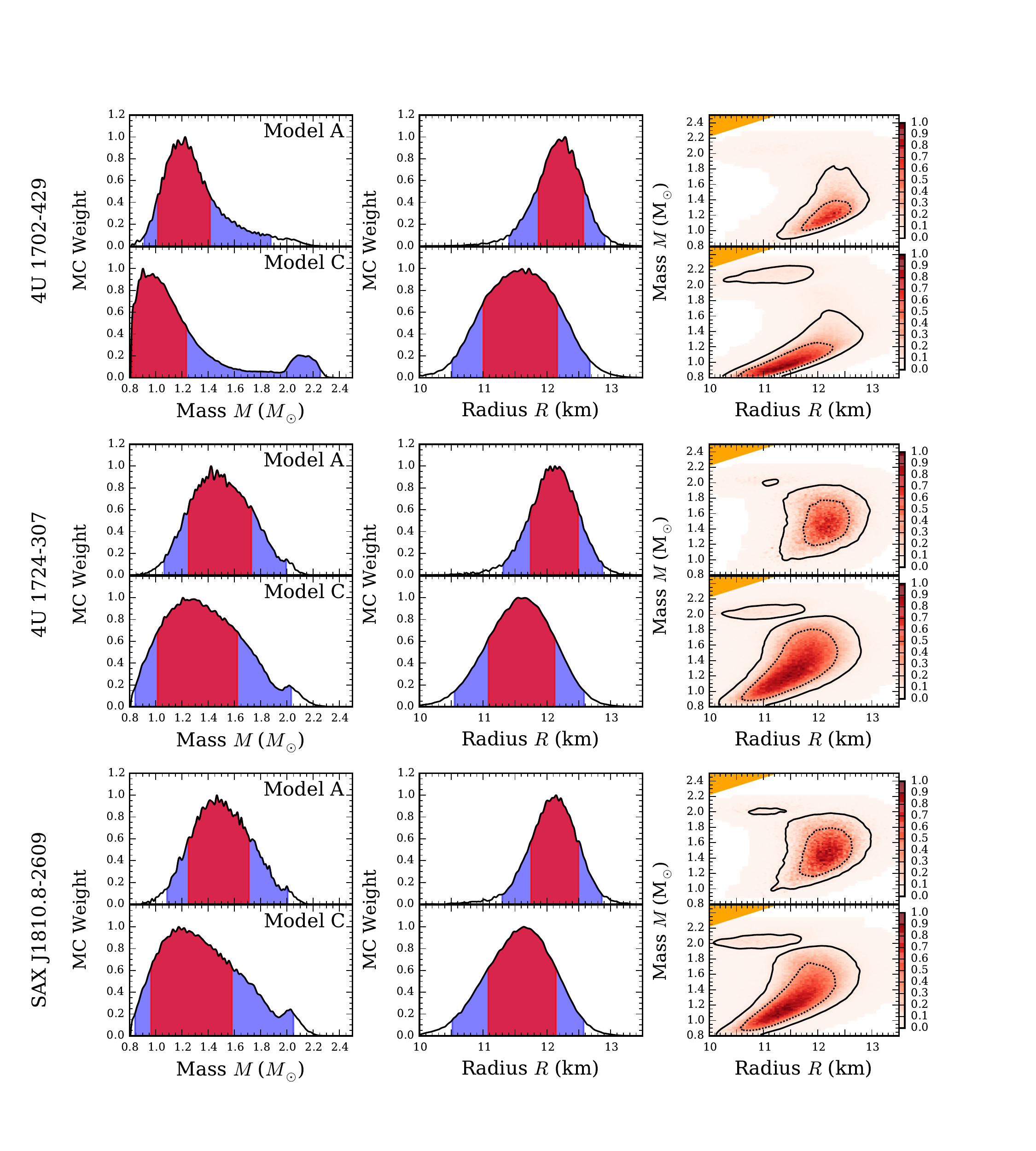}
\caption{\label{fig:mr_sep_Appendix}
  Individual mass and radius constraints for the three neutron stars used in the analysis with uniform $F_{\mathrm{Edd}}$ and $A$ priors.
  Left panel shows the projected mass and the middle panel the projected radius histograms.
  Red shading corresponds to the $68\%$ and blue to the $95\%$ confidence regions of the parameters.
  Right panels shows the full 2-dimensional mass and radius probability distributions.
  Contours of $68\%$ (dotted, black line) and $95\%$ (solid, black line) confidence regions are also shown.
}
\end{figure*}

Secondly, there exists some issues with the transformation from $F_{\mathrm{Edd}}$ and $A$ parameters%
\footnote{
    Our definition of the second $A$ parameter differs from the one used in the work by \cite{OP15} where $\mathcal{A} = R_{\infty}^2 / D^2 f_{\mathrm{c}}^4 = (f_{\mathrm{c}} A)^{-4}$. Conclusions, however, remain the same.
} to $M$ and $R$ as the determinant of the Jacobian is
\be
\det \left\{ J\left(\frac{F_{\mathrm{Edd}}, A}{M, R} \right) \right\} = \frac{c\,G |1-\frac{4GM}{Rc^2}|}{2 D^{3/2} \kappa_{\mathrm{es}} R^{3/2}(1-\frac{2GM}{Rc^2})^{3/2}}.
\ee
From here we can see that we end up ignoring the so-called critical line where $R = 4GM/c^2$, as was shown by \cite{OP15}.
Hence, some information of $M$ and $R$ is already present in $(M,R)$ space owing to our choice of uniform $F_{\mathrm{Edd}}$ and $A$ priors.
The effect from this is visible as a splitting of our $M$ vs. $R$ posteriors into two separate regions (see Fig. \ref{fig:MR_Appendix}).
For clarity, no scatter in $X$ or error in $f_{\mathrm{c}}$ is taken into account in this case.
The lack of $(M,R)$ solutions near this critical line also ends up affecting our final joint-fit results as can be seen from Fig. \ref{fig:mr_sep_Appendix}.
Because of these aforementioned issues we choose uniform priors in $M$ and $R$ instead, as they are our final parameters that we want to study.
We note, however, that in the cases analyzed before by \citet{SPW11} and \citet{PNK14} the error is not critical because the solution on $M-R$ plane is constrained to be far from the $R = 4GM/c^2$ line because of the distance constraints.

\clearpage
\section{EoS Tables and parameter correlations}

\begin{figure*}[!p]
\centering
\includegraphics[width=19cm]{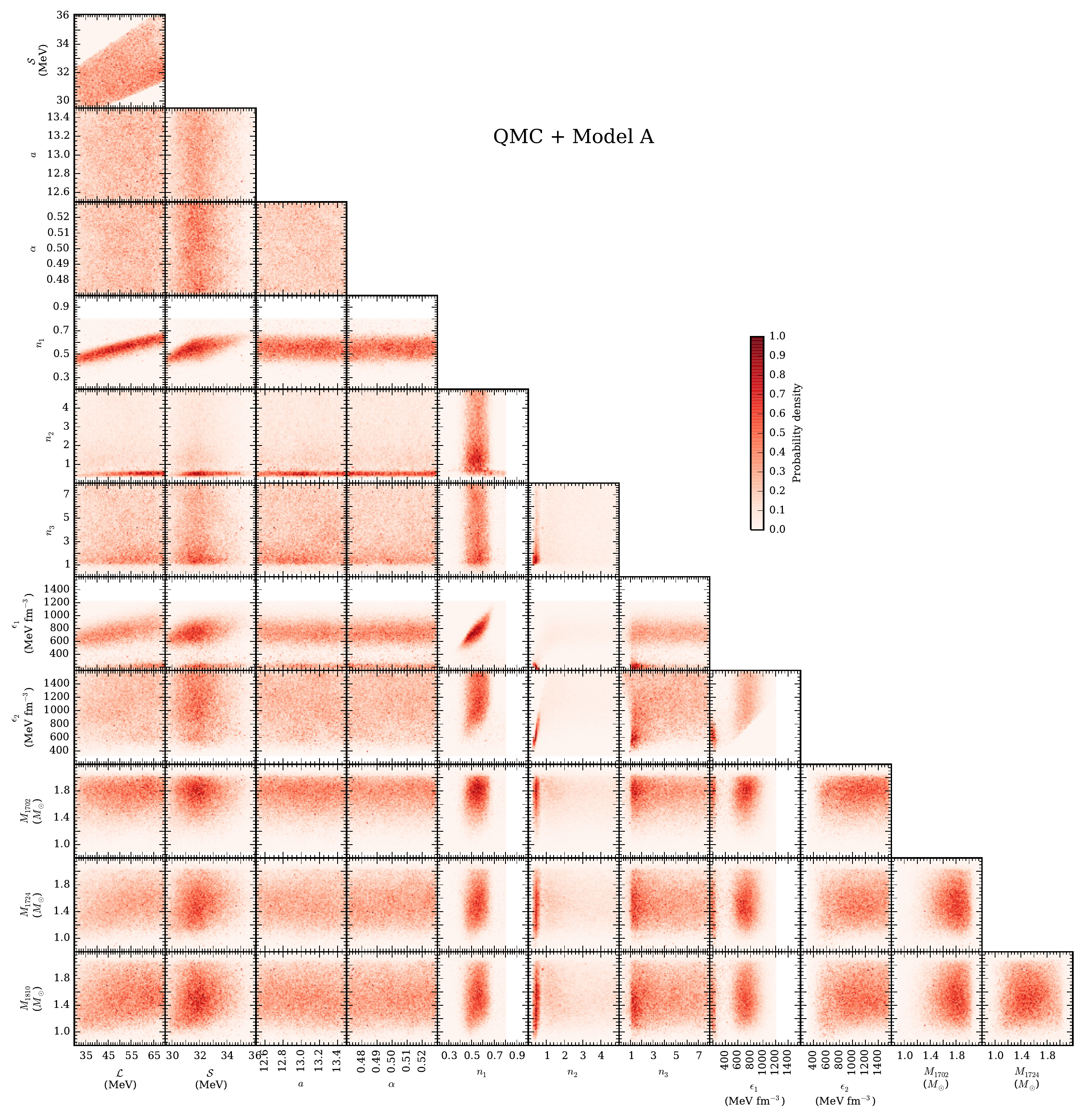}
\caption{\label{fig:corr_A}
  QMC $+$ Model A EoS parameter correlations against each other.
  Red coloring shows the probability density of model parameters obtained from our Bayesian analysis with the cooling tail data.
}
\end{figure*}

\begin{figure*}[!p]
\centering
\includegraphics[width=19cm]{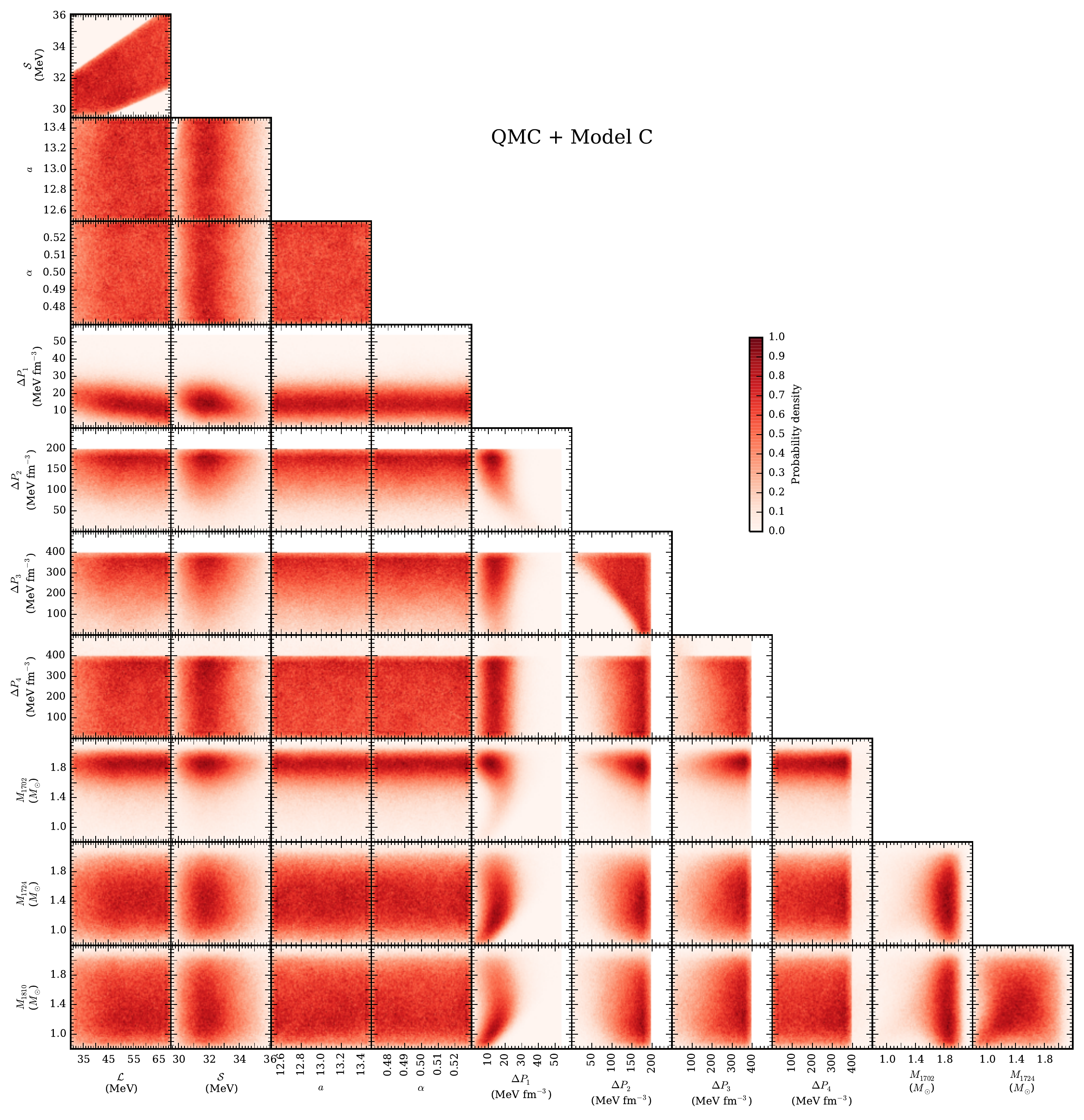}
\caption{\label{fig:corr_C}
  QMC $+$ Model C EoS parameter correlations against each other. Symbols are the same as in Fig. \ref{fig:corr_A}.
}
\end{figure*}

The results for the two possible parameterized EoSs are tabulated here for easier access.
Tables \ref{tab:e_P_A} and \ref{tab:e_P_C} list the pressure vs. density relation for the QMC+Model A and QMC+Model C EoSs, respectively.
Similarly, Tables \ref{tab:m_r_A} and \ref{tab:m_r_C} summarize the resulting mass vs. radius relation for the two aforementioned models.
Additionally, we show all the EoS parameter correlations between each other in Fig.~\ref{fig:corr_A} and \ref{fig:corr_C}.

\clearpage

\begin{table*}
\caption{Most probable values and $68\%$ and $95\%$ confidence limits for the pressure as a function of energy density for QMC $+$ Model A.}
\centering
\begin{footnotesize}  
  \begin{tabular}[c]{l c c c c c}
    \hline
    \hline
  Energy density & $2\sigma$ lower limit & $1\sigma$ lower limit & Most probable value & $1\sigma$ upper limit & $2\sigma$ upper limit \\
  (MeV/fm$^3$)   &                       &                       &  (MeV/fm$^3$)       &                       &                        \\
  \hline
150 & 1.71 & 2.15 & 2.64 & 3.11 & 3.33 \\
200 & 3.8 & 4.74 & 5.73 & 6.6 & 7.18 \\
250 & 6.91 & 9.1 & 10.4 & 12.0 & 13.3 \\
300 & 12.3 & 15.3 & 16.9 & 19.8 & 22.6 \\
350 & 19.7 & 23.6 & 26.0 & 30.4 & 35.9 \\
400 & 29.5 & 34.0 & 36.5 & 44.6 & 55.3 \\
450 & 42.0 & 46.7 & 50.4 & 63.0 & 81.3 \\
500 & 57.2 & 61.9 & 67.8 & 86.1 & 109.9 \\
550 & 74.7 & 80.7 & 86.7 & 113.6 & 139.4 \\
600 & 94.9 & 104.0 & 114.1 & 143.7 & 169.3 \\
650 & 117.7 & 130.6 & 147.6 & 173.4 & 199.2 \\
700 & 143.2 & 157.7 & 171.3 & 202.2 & 229.0 \\
750 & 169.3 & 183.6 & 194.7 & 230.0 & 258.5 \\
800 & 192.9 & 208.6 & 221.3 & 258.5 & 288.7 \\
850 & 213.0 & 233.3 & 251.5 & 288.3 & 319.2 \\
900 & 230.4 & 257.0 & 272.2 & 318.5 & 350.1 \\
950 & 246.2 & 279.1 & 294.3 & 348.2 & 381.2 \\
1000 & 262.3 & 300.2 & 323.7 & 377.2 & 413.9 \\
1050 & 277.6 & 320.8 & 344.4 & 406.3 & 447.0 \\
1100 & 293.2 & 341.3 & 372.6 & 435.7 & 481.8 \\
1150 & 308.7 & 360.8 & 394.7 & 465.0 & 517.9 \\
1200 & 325.1 & 379.9 & 414.8 & 494.6 & 556.1 \\
1250 & 341.0 & 399.0 & 448.8 & 525.0 & 594.5 \\
1300 & 357.7 & 418.0 & 462.0 & 555.8 & 634.5 \\
1350 & 373.2 & 436.7 & 481.1 & 586.8 & 674.8 \\
1400 & 389.2 & 455.6 & 505.5 & 618.8 & 716.4 \\
1450 & 405.1 & 474.5 & 531.2 & 651.7 & 759.3 \\
1500 & 421.4 & 493.0 & 558.2 & 685.1 & 804.6 \\
\hline
 \end{tabular}
\end{footnotesize}
\label{tab:e_P_A}
\end{table*}

\begin{table*}
\caption{Most probable values and $68\%$ and $95\%$ confidence limits for the pressure as a function of energy density for QMC $+$ Model C.}
\centering
\begin{footnotesize}
  \begin{tabular}[c]{l c c c c c}
    \hline
    \hline
  Energy density & $2\sigma$ lower limit & $1\sigma$ lower limit & Most probable value & $1\sigma$ upper limit & $2\sigma$ upper limit \\
  (MeV/fm$^3$)   &                       &                       &  (MeV/fm$^3$)       &                       &                        \\
  \hline
150 & 1.77 & 2.3 & 2.97 & 3.35 & 3.51 \\
200 & 3.19 & 4.38 & 5.46 & 7.32 & 9.03 \\
250 & 3.93 & 6.42 & 8.94 & 12.3 & 15.6 \\
300 & 4.78 & 8.54 & 12.6 & 17.4 & 22.5 \\
350 & 5.65 & 10.7 & 15.1 & 22.6 & 29.3 \\
400 & 11.7 & 17.7 & 22.7 & 31.5 & 38.5 \\
450 & 35.3 & 47.8 & 57.7 & 70.1 & 78.3 \\
500 & 53.1 & 81.1 & 100.8 & 119.0 & 127.6 \\
550 & 70.7 & 114.7 & 145.1 & 168.0 & 177.5 \\
600 & 92.1 & 146.8 & 187.3 & 212.5 & 224.6 \\
650 & 122.7 & 165.9 & 193.1 & 235.8 & 259.8 \\
700 & 156.0 & 181.7 & 198.8 & 259.8 & 300.7 \\
750 & 183.8 & 199.6 & 215.0 & 290.1 & 341.5 \\
800 & 200.5 & 224.7 & 244.4 & 332.2 & 386.9 \\
850 & 212.6 & 254.8 & 277.7 & 381.5 & 436.0 \\
900 & 223.3 & 285.2 & 331.8 & 430.8 & 485.6 \\
950 & 234.3 & 316.8 & 377.0 & 481.3 & 535.7 \\
1000 & 249.7 & 345.3 & 414.7 & 523.1 & 578.9 \\
1050 & 273.4 & 367.9 & 435.7 & 546.9 & 607.1 \\
1100 & 294.9 & 389.4 & 457.9 & 572.5 & 643.0 \\
1150 & 312.7 & 409.4 & 481.1 & 600.4 & 683.6 \\
1200 & 328.1 & 430.1 & 505.5 & 632.4 & 726.4 \\
1250 & 340.8 & 451.1 & 531.2 & 669.2 & 769.6 \\
1300 & 352.4 & 471.7 & 563.2 & 708.8 & 814.1 \\
1350 & 363.1 & 492.0 & 609.2 & 750.4 & 859.0 \\
1400 & 373.1 & 512.4 & 647.6 & 793.8 & 904.5 \\
1450 & 382.6 & 532.8 & 680.4 & 838.4 & 950.5 \\
1500 & 392.3 & 555.0 & 715.0 & 885.7 & 997.0 \\
  \hline
 \end{tabular}
\end{footnotesize}
\label{tab:e_P_C}
\end{table*}

\begin{table*}
\caption{Most probable values and $68\%$ and $95\%$ confidence limits for the NS radii of fixed mass for QMC $+$ Model A.}
\centering
\begin{footnotesize}  
  \begin{tabular}[c]{l c c c c c}
    \hline
    \hline
  Mass & $2\sigma$ lower limit & $1\sigma$ lower limit & Most probable value & $1\sigma$ upper limit & $2\sigma$ upper limit \\
  ($\Msun$)   &                       &                       &  (km)       &                       &                        \\
  \hline
0.5 & 11.60 & 12.05 & 12.79 & 13.20 & 13.73 \\
0.6 & 11.45 & 11.88 & 12.19 & 12.95 & 13.44 \\
0.7 & 11.37 & 11.81 & 12.12 & 12.82 & 13.25 \\
0.8 & 11.34 & 11.78 & 12.12 & 12.73 & 13.13 \\
0.9 & 11.33 & 11.78 & 12.12 & 12.69 & 13.03 \\
1.0 & 11.31 & 11.77 & 12.04 & 12.63 & 12.95 \\
1.1 & 11.32 & 11.75 & 12.04 & 12.57 & 12.90 \\
1.2 & 11.31 & 11.73 & 12.04 & 12.52 & 12.84 \\
1.3 & 11.30 & 11.71 & 12.04 & 12.47 & 12.80 \\
1.4 & 11.27 & 11.67 & 12.04 & 12.42 & 12.75 \\
1.5 & 11.23 & 11.62 & 11.88 & 12.36 & 12.71 \\
1.6 & 11.18 & 11.55 & 11.80 & 12.30 & 12.68 \\
1.7 & 11.10 & 11.45 & 11.72 & 12.22 & 12.63 \\
1.8 & 10.98 & 11.33 & 11.64 & 12.14 & 12.58 \\
1.9 & 10.81 & 11.17 & 11.48 & 12.05 & 12.52 \\
2.0 & 10.49 & 10.93 & 11.40 & 11.95 & 12.46 \\
2.1 & 10.65 & 11.03 & 11.48 & 12.00 & 12.52 \\
2.2 & 10.91 & 11.30 & 11.56 & 12.20 & 12.66 \\
  \hline
 \end{tabular}
\end{footnotesize}
\label{tab:m_r_A}
\end{table*}

\begin{table*}
\caption{Most probable values and $68\%$ and $95\%$ confidence limits for the NS radii of fixed mass for QMC $+$ Model C.}
\centering
\begin{footnotesize}
  \begin{tabular}[c]{l c c c c c}
    \hline
    \hline
  Mass & $2\sigma$ lower limit & $1\sigma$ lower limit & Most probable value & $1\sigma$ upper limit & $2\sigma$ upper limit \\
  ($\Msun$)   &                       &                       &  ($\Msun$)       &                       &                        \\
  \hline
0.5 & 10.17 & 11.61 & 12.76 & 13.63 & 13.87 \\
0.6 & 10.04 & 11.00 & 12.68 & 13.24 & 13.57 \\
0.7 & 10.02 & 10.63 & 11.16 & 12.96 & 13.41 \\
0.8 & 10.05 & 10.38 & 11.06 & 12.28 & 13.30 \\
0.9 & 10.10 & 10.45 & 11.00 & 12.07 & 13.19 \\
1.0 & 10.15 & 10.54 & 11.02 & 11.96 & 13.04 \\
1.1 & 10.22 & 10.63 & 11.08 & 11.90 & 12.84 \\
1.2 & 10.32 & 10.72 & 11.16 & 11.86 & 12.68 \\
1.3 & 10.42 & 10.80 & 11.16 & 11.85 & 12.56 \\
1.4 & 10.52 & 10.88 & 11.24 & 11.85 & 12.47 \\
1.5 & 10.59 & 10.95 & 11.32 & 11.85 & 12.41 \\
1.6 & 10.63 & 11.01 & 11.32 & 11.86 & 12.37 \\
1.7 & 10.63 & 11.06 & 11.40 & 11.88 & 12.34 \\
1.8 & 10.61 & 11.09 & 11.40 & 11.90 & 12.32 \\
1.9 & 10.53 & 11.10 & 11.40 & 11.93 & 12.31 \\
2.0 & 10.38 & 11.06 & 11.45 & 11.98 & 12.31 \\
2.1 & 10.45 & 11.09 & 11.48 & 11.99 & 12.30 \\
2.2 & 10.63 & 11.13 & 11.48 & 11.97 & 12.30 \\
  \hline
 \end{tabular}
\end{footnotesize}
\label{tab:m_r_C}
\end{table*}


\end{appendix}
\end{document}